\documentclass[showpacs,showkeys,preprintnumbers,superscriptaddress,amsmath,floatfix,amssymb,a4paper,secnumarabic]{revtex4}
\usepackage[colorlinks=true]{hyperref}
\usepackage{graphicx}
\usepackage{epsfig}
\usepackage{slashbox}

\newcommand{\beq}{\begin{equation}}
\newcommand{\eeq}{\end{equation}}
\newcommand{\bal}{\begin{aligned}}
\newcommand{\eal}{\end{aligned}}

\def\tr{\hbox{Tr}}
\def\dalam{\hbox
{\vrule\vbox{\hrule\hbox to 1ex{ \hfill}\kern 1 ex\hrule}\vrule}}

\def\1/2{\hbox{$ {1 \over 2}$ }}

\def\h{\hbar}
\def\i/h{{i \over \h}}

\def\arctg{\hbox{arctg}}

\def\inf{\infty}
\def\pd{\partial} 
\def\v{\vec}

\def\a{\alpha} 
\def\b{\beta} 
 
\def\g{\gamma} \def\G{\Gamma} 
\def\d{\delta} \def\D{\Delta}
 
\def\e{\epsilon} \def\E{\hbox{$\cal E $}}

\def\s{\sigma}
\def\r{\rho} 
\def\x{\xi}
 
\def\vf{\varphi}

\def\p{\psi}

\def\m{\mu}
\def\n{\nu}

\def\tt{\theta}

\def\<{\langle}
\def\>{\rangle}

\def\({\left(}
\def\[{\left[}
\def\){\right)}
\def\]{\right]}

\def\cM{\mathcal{M}}
\def\cW{\mathcal{W}}
\def\cK{\mathcal{K}}
\def\cI{\mathcal{I}}

\begin{document}

\title{Non-perturbative vacuum polarization effects in two-dimensional supercritical Dirac-Coulomb system. I. Vacuum charge density}

\author{A.~Davydov}
\email{davydov.andrey@physics.msu.ru} \affiliation{Department of Physics and
Institute of Theoretical Problems of MicroWorld, Moscow State
University, 119991, Leninsky Gory, Moscow, Russia}

\author{K.~Sveshnikov}
\email{costa@bog.msu.ru} \affiliation{Department of Physics and
Institute of Theoretical Problems of MicroWorld, Moscow State
University, 119991, Leninsky Gory, Moscow, Russia}

\author{Y.~Voronina}
\email{voroninayu@physics.msu.ru} \affiliation{Department of Physics and
Institute of Theoretical Problems of MicroWorld, Moscow State
University, 119991, Leninsky Gory, Moscow, Russia}

\begin{abstract}
 Based on the original combination of analytical methods, computer algebra tools and numerical calculations, proposed recently in Refs.~\cite{davydov2017}-\nocite{sveshnikov2017}\cite{voronina2017}, the non-perturbative vacuum polarization effects    in the 2+1 D supercritical Dirac-Coulomb system with $Z > Z_{cr,1}$ are explored. Both the  vacuum charge density $\r_{VP}(\v r)$ and vacuum energy $\E_{VP}$ are considered. The main result of the work is that in the overcritical region $\E_{VP}$  turns out  to be a rapidly decreasing function  $\sim - \eta_{eff}\, Z^3/R\, $ with $\eta_{eff}>0 $ and $R$ being the size of the external Coulomb source. Due to a lot of details of calculation the whole work is divided into two parts I and II. In the present  part I we consider the evaluation and behavior of the vacuum density $\r_{VP}$, which further is used in the part II for evaluation of the vacuum energy, with emphasis on the  renormalization, convergence of the partial expansion for $\r_{VP}$ and behavior of the integral induced charge  $Q_{VP}$ in the overcritical region.
\end{abstract}

\maketitle
%%pacs  numbers should be changed - see http://publish.aps.org/PACS

\section{Introduction}

There is now a lot of interest to the non-perturbative effects in the QED vacuum, caused by diving of discrete levels into the lower continuum in the  supercritical static or adiabatically slowly varying Coulomb fields, that are created by localized extended sources with $Z> Z_{cr}$ (see Refs.~\cite{raf2017}-\nocite{NP2015, popov2015, superheavy2013}\cite{ruffini2010} and refs. therein). Such effects have  attracted a considerable amount of theoretical and experimental activity, since in 3+1 QED for $Z > Z_{cr,1} \simeq 170$ a non-perturbative reconstruction of the vacuum state is predicted, which should be accompanied by a number of nontrivial effects including the vacuum positron emission (see Refs.~\cite{raf2017},\cite{rg1977}-\nocite{greiner1985, plunien1986}\cite{greiner2012} and refs. therein). Similar in essence effects are expected to come out also  both in  2+1 D (planar graphene-based hetero-structures \cite{katsnelson2006,shytov2007,nomura2007,kotov2008,pereira2008,herbut2010,wang2013,nishida2014}) and  in 1+1 D (one-dimensional ``hydrogen ion''~\cite{barb1971,krainov1973,shabad2006,shabad20061,shabad2008,oraevskii1977,karn2003,vys2014}). Moreover, in the graphene case the role of the effective QED vacuum  is played by the graphene itself, the electrons and holes are the analogues of virtual pairs, while the effective fine-structure coupling constant  $\a_g \sim 1$.

Recently, in Refs.~\cite{davydov2017}-\nocite{sveshnikov2017}\cite{voronina2017} an original combination of analytical methods, computer algebra tools and numerical calculations has shown that for a wide range  of the system parameters in the one-dimensional Dirac-Coulomb problem the nonlinear effects  could lead in the supercritical region to the behavior of the vacuum energy, substantially different from the perturbative quadratic growth,  up to (almost) quadratic decrease into the negative region $\sim -|\eta| Z^2$. In the present paper, which is divided into two parts  I and II,  these methods are applied to the study of analogous vacuum polarization effects  for a  2+1  Dirac-Coulomb system in the overcritical region. More concretely, in the present  part I we consider the behavior of the vacuum charge density $\r_{VP}$, while in  the subsequent part II with account of results, obtained in I, the behavior of the vacuum energy $\E_{VP}$ is explored.

The external Coulomb field $A^{ext}_{0}(\v r)$ is chosen in the form of a projection onto a plane of the potential of the uniformly charged sphere with radius  $R$
\beq \label{1.00}
A^{ext}_{0}(\v r)=Z |e| \[\frac{1}{R}\tt\(R-r\)+\frac{1}{r}\tt\(r-R\) \] \ ,
\eeq
what leads to the potential energy
\beq
\label{1.1}
V(r)= - Z \a \[\frac{1}{R}\tt\(R-r\)+\frac{1}{r}\tt\(r-R\) \] \ .
\eeq
Compared to the model of the uniformly charged ball such potential turns out to be more preferable, since it allows to perform the most part of calculations in the analytical form, while the evaluation of critical charges shows that  in both cases the final answers should be quite close. And although with the standard choice of the fine-structure coupling $\a \simeq 1/137$ and without special selection of the Coulomb field cut-off parameters such a system could be treated only as a toy-model of 3+1 D problem,  its study for $Z > Z_{cr}$ should be of considerable interest, since it allows to reproduce almost all the properties of the realistic 3+1 D problem of vacuum polarization by superheavy nuclei or nuclear quasi-molecule, but with certain substantial simplifications due to  a smaller number of rotational degrees of freedom.  For these reasons the radius of the external source  is chosen as in 3+1 D for the case of superheavy nuclei
\beq
\label{1.2}
R=R(Z) \simeq 1.2\, (2.5\, Z)^{1/3} \ \text{fm} \ .
\eeq

As it was shown in Refs.~\cite{davydov2017}-\nocite{sveshnikov2017}\cite{voronina2017},  the actual calculation of the vacuum energy, based on the  UV-renormalization  of the fermionic loop, could be performed without resorting  to the  vacuum charge density and shell effects.  However, in fact the decrease of  $\E_{VP}^{ren}$ in the overcritical region is caused primarily by the non-perturbative changes in the vacuum density for $Z > Z_{cr,1}$ due to the discrete levels diving into the lower continuum (``the shell effect'').  In 1+1 D, due to the specifics of one-dimensional Dirac-Coulomb problem, the total vacuum shells number depends on $Z$ as  $\sim Z^s \ , \ 1<s<2$, at least within the considered in Refs.~\cite{davydov2017}-\nocite{sveshnikov2017}\cite{voronina2017} range of external parameters. Therefore, they turn out to be able just to reduce the growth rate of the non-renormalized  $\E_{VP}$ in the overcritical region up to $\sim Z^\n$, $1<\n<2$, and so  the dominant contribution comes from the renormalization term  $\eta Z^2$. In 2+1 and 3+1 D the shell effect shows up much more pronouncedly. As a consequence, in this case $\E_{VP}^{ren}(Z)$ behaves in the overcritical region remarkably more nonlinearly, what will be demonstrated in the part II of the present work for the same 2+1 Dirac-Coulomb system  with the external field (\ref{1.00}).

It is also worth-while noticing that our work is aimed mainly at the study of vacuum effects, caused by extended supercritical Coulomb sources with non-zero size $R$ like superheavy nuclei or charged impurities in graphene, which provide a physically clear problem statement. At the same time, certain  aspects of the physics, associated with supercritical point-like charges, have been studied by many authors for a long time. In particular, in Ref.~\cite{gartner1981} this phenomenon
was investigated with regard to the concept of
the vacuum charge. It turned out that when the radius of a
supercritical nucleus ($Z > 137$) tends to zero, the vacuum
charge screens the nuclear charge to $137$ that prevents a
further diving of the electron states. This result leads to
the conclusion that the interaction with a point charge
in QED cannot effectively have the
coupling strength greater than $1$~\cite{greiner1985,gartner1981} . In Ref.~\cite{aleksandrov2016} it was explored, how the previous statement
may alter in the presence of nuclear recoil and vacuum polarization
operators. Purely mathematical aspects of supercritical point-like charges have been also extensively discussed in terms of possible self-adjoint extensions of the Dirac hamiltonian. A quite  rigorous and comprehensive treatment of this problem, based on the theory of self-adjoint extensions of symmetric operators combined with  the Krein's method of directional functionals,  was presented in
Refs.~\cite{voronov2007},\cite{gitman2013} (see also refs. therein). However, despite the fact that a rigorous
and consistent theory of self-adjoint operators was applied
to the problem, the electronic states cannot be still completely
determined for $Z > 137$. There remains an ambiguity
      in choosing the distinguished one between the possible self-adjoint extensions. Our work deals with extended sources only and so is free from such ambiguities.

As in other works on vacuum polarization in the strong  Coulomb field, radiative corrections from virtual photons are neglected. Henceforth, if it is not stipulated separately, relativistic units  $\hbar=m_e=c=1$ are used. Thence the coupling constant $\a=e^2$ is also dimensionless, what significantly simplifies the subsequent analysis, while the numerical calculations, illustrating the general picture, are performed for  $\a=1/137.036$.

\section{Perturbation Theory for the Vacuum Density in 2+1 QED}

In the 2+1 QED within the perturbation theory (PT) the leading order  vacuum charge density $\r^{(1)}_{VP}(\vec{r})$ is defined by means of the relation
\beq
\label{2.2}
\r^{(1)}_{VP}(\vec{r})=-\frac{1}{4 \pi} \D_{2}\, A^{(1)}_{VP,0}(\vec{r}) \ ,
\eeq
where $\D_2$ is the two-dimensional Laplace operator. In (\ref{2.2}) the Uehling potential $A^{(1)}_{VP,0}(\v r)$ is expressed via the polarization operator $\Pi_R(-\vec{q}\,^2)$ and the Fouriet-transform of the external potential  $\widetilde{A}_{0}(\vec{q})$ \cite{greiner2012}
\beq
\label{2.3}
\begin{aligned}
&A^{(1)}_{VP,0}(\vec{r})=\frac{1}{(2 \pi)^2} \int d^2q\,\mathrm{e}^{i \vec{q} \vec{r}} \Pi_{R}(-q^2)\widetilde{A}_{0}(\vec{q}) \ , \\
&\widetilde{A}_{0}(\vec{q})=\int d^2 r'\,\mathrm{e}^{-i \vec{q} \vec{r\,}' }A^{ext}_{0}(\vec{r}\,' ) \ , \qquad q=|\vec{q}| \ ,
\end{aligned}
\eeq
where
\beq
\label{2.4}
\Pi_R(-q^2)=\frac{\a}{2q} \[\frac{2 }{q} + \(1-\frac{4}{q^2}\) \arctg \( \frac{q}{2}\)\] \ .
\eeq
Here the polarization operator  is defined through relation  $\Pi_R^{\m\n}(q)=\(q^\m q^\n - g^{\m\n}q^2\)\Pi_R(q^2)$ and so is dimensionless. Calculation of the explicit form of the polarization function (\ref{2.4}) is based on the two-dimensional representation of Dirac matrices (concerning the latter choice see below in Section 3).
From (\ref{2.3}), (\ref{2.4}) for the external source (\ref{1.00}) there follows the explicit expression for the corresponding Uehling potential in the form of an axial-symmetric function (the details of calculations are given in Appendix A)
\beq
\label{2.5}
\bal
A_{VP,0}^{(1)}(r)&=\frac{Z \a |e|}{4}\int\limits_{0}^{\infty}dq\,\frac{J_0(qr)}{q}\[\frac{2 }{q} + \(1-\frac{4}{q^2}\) \arctg \( \frac{q}{2}\)\] \times \\\
&\times \(2\[1+J_1(q R)-q R J_0(q R)\] +\pi q R\[ J_0(q R) \mathbf{H}_1(q R)- J_1(q R) \mathbf{H}_0(q R)\]\) \ ,
\eal
\eeq
with  $J_{\n}(z)$ being the Bessel function and $\mathbf{H}_{\n}(z)$ --- the Struve one.

Proceeding further, in  the next step by means of (\ref{2.2}) one  finds the axial-symmetric vacuum charge density $\r^{(1)}_{VP}(r)$. To clarify the question of the possibility of inserting the Laplace operator under the sign of the integral over $d q$ in (\ref{2.5}), let us consider the asymptotics of the integrand for large $q$. The leading term of the asymptotics takes the form
\beq \label{2.5a}
\frac{ \sin (q (r+R))+\cos (q (r-R))}{\sqrt{r}R^{3/2}\,q^{3}} \ .
\eeq
The action of the Laplace operator in (\ref{2.5a}) yields the term, which behaves   $\sim 1/q$ at $r=R$, namely
\beq \label{2.5b}
\D_{2} \frac{\sin (q (r+R))+\cos (q (r-R))}{\sqrt{r}R^{3/2}\,q^{3}}= - \frac{\sin (q (r+R))+\cos (q (r-R))}{\sqrt{r}R^{3/2}\,q} + \mathrm{O}(1/q^2).
\eeq
Therefore, at  $r=R$ the possibility of  inserting the Laplace operator under the sign of the integral in (\ref{2.5}) is absent, since in this case, according to (\ref{2.5b}), the integral over $d q$ diverges logarithmically.

Let us also mention that the  factor $1/\sqrt{r}$ in  (\ref{2.5b}) does not produce any singularity in the behavior of $A^{(1)}_{VP,0}(r)$ for $r \to 0$, since the asymptotics (\ref{2.5a}) of the integrand in (\ref{2.5}) is achieved by taking the limit $ q r \to \inf$ in the argument of the Bessel function  $J_0(qr)$. The correct behavior of $A^{(1)}_{VP,0}(r)$ for $r \to 0$ should be found from (\ref{2.5}) by taking $q r \to 0$ in $J_0(qr)$, which leads to the converging integral over $d q$, hence, $A^{(1)}_{VP,0}(0)$, as well as $\r^{(1)}_{VP}(0)$, is finite.

%It would be also worthwhile noticing that the asymptotics (\ref{2.5b}) cannot be used in the infinitesimal vicinity of the point $r=0$, since  $r$ enters in $A_{VP,0}^{(1)}$ (\ref{2.5}) via combination $q r$, which for $q \to +\inf$ in the vicinity of $r=0$ might remain finite. From (\ref{2.5c}), one can easily see that $\r_{VP}^{(1)}$ is finite at $r=0$.

So the vacuum density, obtained from (\ref{2.2}) with account of (\ref{2.5b}), equals to
\beq
\label{2.5c}
\bal
\r_{VP}^{(1)}(r)&=\frac{Z \a |e|}{16\pi}\int\limits_{0}^{\infty}dq\,q J_0(q r)\[\frac{2 }{q} + \(1-\frac{4}{q^2}\) \arctg \( \frac{q}{2}\)\] \times \\\
&\times \(2\[1+J_1(q R)-q R J_0(q R)\] +\pi q R\[ J_0(q R) \mathbf{H}_1(q R)- J_1(q R) \mathbf{H}_0(q R)\]\)
\eal
\eeq
and is finite for all  $r\neq R$ with the logarithmic singularity at $r \to  R$.

 From (\ref{2.5c}) by an explicit calculation one obtains that to the leading order of PT the integral vacuum charge  vanishes exactly
\beq
\label{2.7}
\int \! d^2r \ \r^{(1)}_{VP}(r)= 0 \ .
\eeq
And although in this case the relation (\ref{2.7}) turns out to be the direct consequence of the renormalization condition ${\tilde  \Pi_R} (q^2) \sim  q^2$ for $q \to 0$ (see Appendix B),
actually it should be considered as an additional indication in favor of the assumption, that in presence of the external field, uniformly vanishing at the spatial infinity, and without any special boundary conditions and/or nontrivial topology of the field manifold, one should expect that in the undercritical region with $Z<Z_{cr,1}$ the correctly renormalized integral vacuum charge should vanish, while the vacuum polarization could only distort its spatial distribution \cite{greiner2012,mohr1998}. It should be noted, however, that this is not a theorem, but just a plausible statement, which in any concrete case should be verified via direct calculation. In the case under consideration the direct check shows (see Appendix B) that upon renormalization the induced vacuum charge turns out to be  non-vanishing only for $Z>Z_{cr,1}$ due to the non-perturbative effects, caused by diving of discrete levels into the lower continuum in accordance with Refs.~\cite{raf2017},\cite{rg1977}-\cite{greiner2012}. And in the part II it will be shown, how the latter circumstance shows up in the behavior of the vacuum energy in the overcritical region.

\section{The Wichmann-Kroll Approach for 2+1 QED}

The most efficient non-perturbative approach to evaluation of the vacuum density $\r_{VP}(\vec{r})$ is based on the method of Wichmann and Kroll (WK) \cite{wk1956} (see also Ref.~\cite{mohr1998} and refs. therein).
The starting point of the WK approach is the following expression for the induced charge density
\beq
\label{3.1}
\r_{VP}(\vec{r})=-\frac{|e|}{2}\(\sum\limits_{\e_{n}<\e_{F}} \p_{n}(\vec{r})^{\dagger}\p_{n}(\vec{r})-\sum\limits_{\e_{n}\geq \e_{F}} \p_{n}(\vec{r})^{\dagger}\p_{n}(\vec{r}) \),
\end{equation}
where  in such problems with external Coulomb source $\e_F$ should be chosen at the threshold of the lower continuum, i.e. $\e_F=-1$, while $\e_{n}$ and $\p_n(\vec{r})$  are the eigenvalues and eigenfunctions of the corresponding  Dirac-Coulomb spectral problem (DC).

The essence  of the WK-method is that the vacuum density
(\ref{3.1}) can be represented via contour integration of the trace of the Green function of the spectral DC-problem in the complex energy plane. By definition, the Green function satisfies the equation
\beq
\label{3.2}
\(- i \,\vec{\a}\,\vec{\nabla}+V(\vec{r})+\b -\e \)G(\vec{r},\vec{r}\,' ;\e)=\d(\vec{r}-\vec{r}\,' ) \ .
\eeq
Here it should be noted that in 2+1 D there are two types of spinors, responsible for two possible choices of the signature of two-dimensional Dirac matrices  \cite{hosotani1993,halilov2016}. At the same time, in 2+1 D the Dirac matrices could also be  taken as four-dimensional ones. In the latter case, the DC spectral problem in the external field (\ref{1.00}) for the four-component wavefunction splits into two independent subsystems, which transform into each other under the change of the sign of the total angular momentum $m_j \to -m_j$. Hence, the degeneracy of states with definite value of $m_j$ equals to 2, and in what follows this factor will be explicitly shown in all the expressions for $\E_{VP}$ and $\r_{VP}$, while the DC spectral problem without any loss of generality will be  considered within the two-dimensional representation with  $\a_i=\s_i$, $\b=\s_3$.

The formal solution of (\ref{3.2}) reads
\beq
\label{3.3}
G(\vec{r},\vec{r}\ ';\e)=\sum\limits_{n}\frac{\p_{n}(\vec{r})\p_{n}(\vec{r}\ ')^{\dagger}}{\e_{n}-\e} \ .
\eeq
Following Ref.~\cite{wk1956}, in the next step the vacuum charge density is expressed via the integrals along the contours  $P(R_0)$ and $E(R_0)$ on the first sheet of the Riemann energy surface  (Fig.1)
\beq
\label{3.4}
\r_{VP}(\vec{r})=-\frac{|e|}{2} \lim_{R_0\rightarrow \infty}\( \frac{1}{2\pi i}\int\limits_{P(R_0)}d\e\, \mathrm{Tr}G(\vec{r},\vec{r};\e)+\frac{1}{2\pi i}\int\limits_{E(R_0)}d\e\, \mathrm{Tr}G(\vec{r},\vec{r};\e) \) \ .
\eeq
\begin{center}
\includegraphics[scale=0.22]{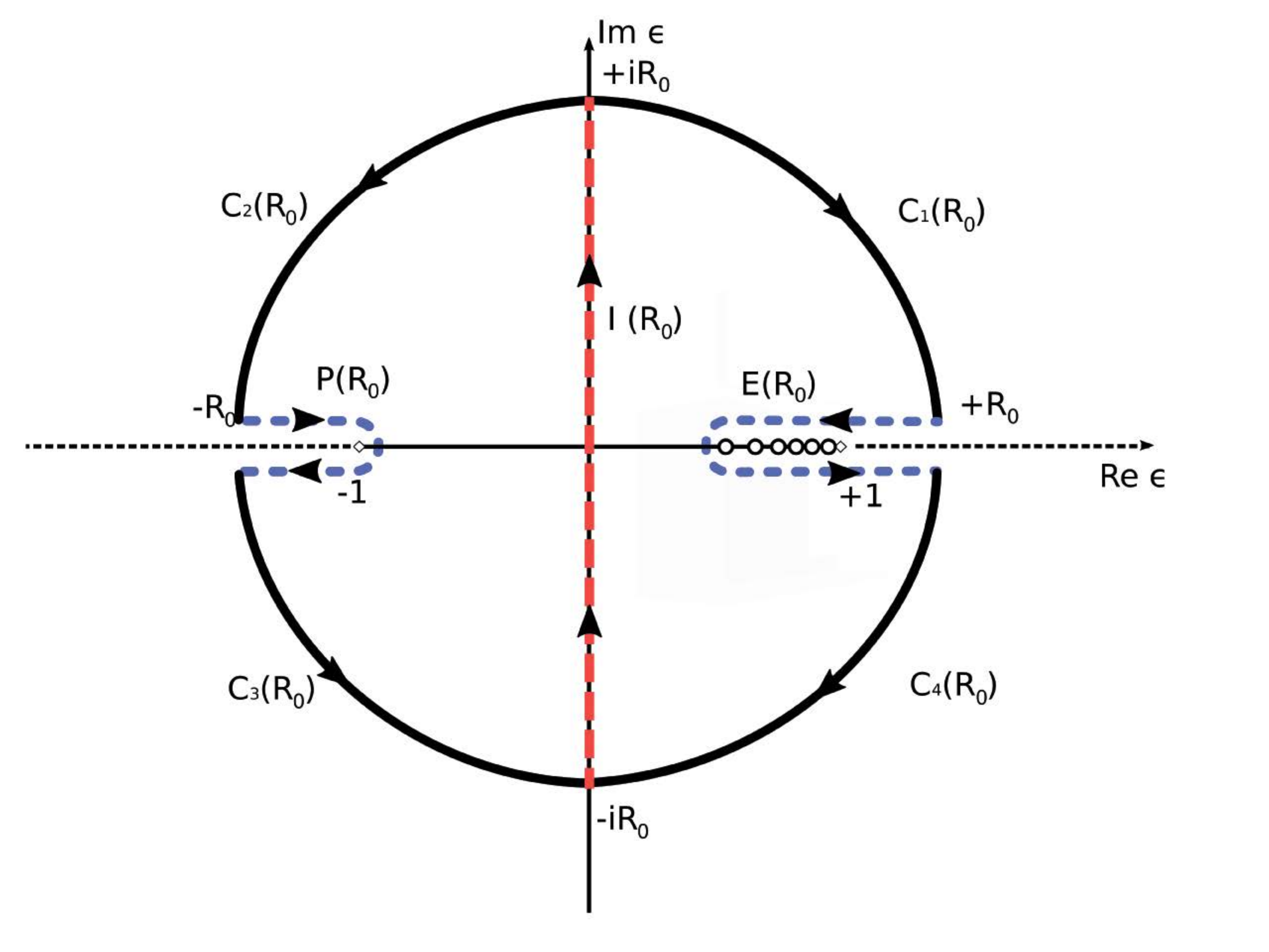} \\
{\small Fig.1. (Color online) Special contours in the complex energy plane, used for representation of the vacuum charge density via contour integrals. The direction of contour integration is chosen in correspondence with (\ref{3.3}).}
\end{center}
Taking account of that the DC spectral problem in the external field (\ref{1.00}) allows for separation of radial and angular variables in the form
\beq
\p(\vec{r})={1 \over \sqrt{2\pi}} \begin{pmatrix}
i \p_{1}(r)\mathrm{e}^{i(m_j-1/2)\vf}\\
\p_{2}(r)\mathrm{e}^{i(m_j+1/2)\vf}
\end{pmatrix} \ ,
\eeq
where $m_j$ is the half-integer total angular momentum, one finds for the trace of the Green function (\ref{3.3})
\beq
\label{3.5}
\begin{aligned}
&\mathrm{Tr}G(\vec{r},\vec{r};\e)\equiv {1 \over 2\pi}\mathrm{Tr}G(r,r;\e)={2 \over 2\pi}\,\sum\limits_{m_j=\pm 1/2, \pm 3/2,..}\mathrm{Tr}G_{m_j}(r,r;\e) \ , \\
&\mathrm{Tr}G_{m_j}(r,r;\e)={\p^{in}_{m_j}(r;\e)^{\mathrm{T}}\p^{out}_{m_j}(r;\e) \over J_{m_j}(\e)} \ ,
\end{aligned}
\eeq
where $\p^{in}_{m_j}(r;\e)$ and $\p^{out}_{m_j}(r;\e)$ are the solutions of the radial spectral DC problem  with the same $m_j$, which are regular for $r \to 0$ and  $r \to +\infty$ correspondingly, while $J_{m_j}(\e)$ is their Wronskian
\beq
\label{3.6}
J_{m_j}(\e)=[\p^{in}_{m_j}(r;\e),\p^{out}_{m_j}(r;\e)] \ .
\eeq
In (\ref{3.6}), a convenient for what follows denotation
\begin{equation*}
\[f,g\]_{a}=a(f_{2}(a)g_{1}(a)-f_{1}(a)g_{2}(a))
\end{equation*}
is introduced.

Defined in such a way, $\tr G_{m_j}$ possesses the required normalization.
It should be noted also that actually $J_{m_j}(\e)$ is nothing else, but the Jost function of the  DC problem for the given $m_j$: the real-valued zeros of $J_{m_j}(\e)$ lie on the first sheet in the interval $-1 \leq \e <1$ and coincide with discrete  levels $\e_{n,m_j}$, while the complex ones reside on the second sheet with negative imaginary part of the wavenumber $k=\sqrt{\e^2-1}$, and for $\mathrm{Re}\, k >0$ correspond to elastic resonances.

For the external potential (\ref{1.00}) $\tr G_{m_j}$  can be easily found in the analytical form. For the fixed $m_j$, the radial DC spectral problem takes the form of the system
\beq
\label{3.7}
\left\lbrace\bal
&\frac{d}{d r}\p_1(r)+\frac{1/2-m_j}{r}\,\p_1(r)=(\e-V(r)+1)\p_2(r) \ ,\\
&\frac{d}{d r}\p_2(r)+\frac{1/2+m_j}{r}\,\p_2(r)=-(\e-V(r)-1)\p_1(r) \ .
\eal\right.
\eeq
For $r\leq R$ it is convenient to choose the linearly independent solutions of the system (\ref{3.7}) as the following ones
\beq
\label{3.8}
\bal
\text{for $\p_1(r;\e)$}: \
&\mathcal{I}_1(r)=\x I_{|m_j-1/2|}(\x r) \ , \  &\mathcal{K}_1(r)&=-\x K_{|m_j-1/2|}( \x r) \ ;\\
\text{for $\p_2(r;\e)$}:\  &\mathcal{I}_2(r)=\(1-\e-V_0\) I_{|m_j+1/2|}(\x r) \ , &\mathcal{K}_2(r)&=\(1-\e-V_0\) K_{|m_j+1/2|}(\x r) \ ;
\eal
\eeq
with $I_{\nu}(z)$ and $K_{\nu}(z)$ being the Infeld and Macdonald functions correspondingly,
\beq
V_0=Z \a/R,\quad \x=\sqrt{1-(\e+V_0)^2}, \quad \mathrm{Re} \, \x\geq 0 \ .
\eeq
For $r>R$ the fundamental pair of solutions of (\ref{3.7}) should be taken in the form
\beq
\label{3.9}
\begin{aligned}
\text{for $\p_1(r;\e)$}: \quad	&\cM_{1}(r)=\frac{1+\e}{r}\[\(s-\n\)M_{\n-1/2,s}(2\g r) +\(m_j+\frac{Q}{\g}\)M_{\n+1/2,s}(2\g r)\] \ ,\\
&\cW_{1}(r)=\frac{1+\e}{r}\[\(m_j-\frac{Q}{\g}\)W_{\n-1/2,s}(2\g r) -W_{\n+1/2,s}(2\g r)\] \ ;\\
\text{for $\p_2(r;\e)$}:\quad	&\cM_{2}(r)=\frac{\g}{r}\[\(s-\n\)M_{\n-1/2,s}(2\g r) -\(m_j+\frac{Q}{\g}\)M_{\n+1/2,s}(2\g r)\] \ ,\\
	&\cW_{2}(r)=\frac{\g}{r}\[\(m_j-\frac{Q}{\g}\)W_{\n-1/2,s}(2\g r) + W_{\n+1/2,s}(2\g r)\] \ ;
\end{aligned}
\end{equation}
with $M_{b,c}(z)$ , $W_{b,c}(z)$ being the Whittacker functions \cite{bateman1953},
\beq
\label{3.10}
\begin{aligned}
	\quad Q=Z \a \ ,  \quad s=\sqrt{m_j^2-Q^2} \ ,\quad \n=\frac{\e\, Q}{\g} \ ,\quad \g=\sqrt{1-\e^2} \ ,  \quad \mathrm{Re}\,\g\geq0 \ .
\end{aligned}
\eeq
The functions $\p^{in}_{m_j}(r;\e)$ and $\p^{out}_{m_j}(r;\e)$, which enter into $\tr G_{m_j}$, are constructed now as such linear combinations of solutions (\ref{3.8}) and (\ref{3.9}), that satisfy the conditions of regularity for $r \to 0$ and $r \to +\infty$ correspondingly. As a result, the final expression for  $\tr G_{m_j}$ takes the form
\beq \label{3.11}
\tr G_{m_j}(r,r;\e)=\left\lbrace
\bal
&\frac{1}{\[\cI,\cK\]}\(\cI_{1}\cK_{1}+\cI_{2}\cK_{2}-\frac{\[\cK,\cW\]_{R}}{\[\cI,\cW\]_{R}}\(\cI_{1}^{2}+\cI_{2}^{2}\)\) \ ,  \ \ &r\leq R \ ,\\
&\frac{1}{\[\cM,\cW\]}\(\cM_{1}\cW_{1}+\cM_{2}\cW_{2}-\frac{\[\cI,\cM\]_{R}}{\[\cI,\cW\]_{R}}\(\cW_{1}^{2}+\cW_{2}^{2}\)\) \ ,  &r > R \ ,
\eal\right.
\eeq
where
\beq \label{3.13}
[\cI,\cK]=\e+V_{0}-1 \ ,  \qquad \[\cM,\cW\]= -4\g^2 (1+\e)\frac{\G(2s+1)}{\G(s-\n)} \ ,
\eeq
while the Wronskian, which enters into the expression for $\tr G_{m_j}$ (\ref{3.5}), equals to
\beq
\label{3.14}
J_{m_j}(\e)=\[\cI,\cW\]_{R} \ .
\eeq

Proceeding further, in the next step one finds the asymptotics of $\tr G_{m_j}$  on the arcs  $C_1(R_0)$ and $C_2(R_0)$ in the upper half-plane (Fig.1) for $|\e|\to \inf$, $0<\mathrm{Arg}\, \e <\pi$:
\beq
\label{3.16}
\bal
&\tr G_{m_j}(r,r;\e)\to \\
&\left\lbrace
\bal
&\frac{i}{r}+\frac{i}{2r \e^2} \(\frac{m_j^2}{r ^2}+1\)-\frac{i}{r^2 \epsilon ^3} \(\frac{m_j^2}{r} V_{0}+\frac{m_j}{2 r}+r V_{0}\)+\mathrm{O}\(|\e|^{-4}\),\quad &r< R,\\
&\frac{i}{r}+\frac{i}{2r \e^2} \(\frac{m_j^2}{r ^2}+1\)-\frac{i}{r ^2 \epsilon ^3} \(\frac{m_j^2 }{r}\frac{Q}{r}+\frac{m_j}{2r}+Q\)+\mathrm{O}\(|\e|^{-4}\),\quad &r>R,
\eal
\right.
\eal
\eeq
and on the arcs  $C_3(R_0)$ and $C_4(R_0)$ in the lower half-plane $|\e|\to \inf$,  $-\pi<\mathrm{Arg}\, \e <0$:
\beq
\label{3.17}
\bal
&\tr G_{m_j}(r,r;\e)\to \\
&\left\lbrace
\bal
&-\frac{i}{r}-\frac{i}{2r \e^2} \(\frac{m_j^2}{r ^2}+1\)+\frac{i}{r ^2 \e ^3} \(\frac{m_j^2}{r} V_{0}+\frac{m_j}{2 r}+r V_{0}\)+\mathrm{O}\(|\e|^{-4}\),\quad &r< R,\\
&-\frac{i}{r}-\frac{i}{2r \e^2} \(\frac{m_j^2}{r ^2}+1\)+\frac{i}{r ^2 \epsilon ^3} \(\frac{m_j^2}{r}\frac{Q}{r}+\frac{m_j}{2 r}+Q\)+\mathrm{O}\(|\e|^{-4}\),\quad &r>R.
\eal
\right.
\eal
\eeq
Upon integration over the arcs, the leading terms in the asymptotics (\ref{3.16},\ref{3.17}) give $(i/r)(-2i\, R_0)$ for the contribution from $C_1+C_2$ and $(i/r)(+2i\, R_0)$ from $C_3+C_4$, which cancel each other. At the same time, the integration of the next-to-leading terms in (\ref{3.16},\ref{3.17}) contains always inverse powers of $R_0$ in the answer, hence, their contribution vanishes for $R_0 \to \inf$.

So there follows from (\ref{3.16}) and (\ref{3.17}), that the integration along the contours $P(R_0)$ and $E(R_0)$ (Fig.1) in (\ref{3.4}) could be reduced to the imaginary axis, whence one obtains the final expression for the vacuum charge density
\beq
\label{3.18}
\r_{VP}(r)=2 \sum\limits_{m_j=1/2,\,3/2,..} \r_{VP,|m_j|}(r)\ ,
\eeq
where
\beq \label{3.18a}
\bal
&\r_{VP,|m_j|}(r)=\frac{|e|}{(2 \pi)^2}\int\limits_{-\inf}^{\inf}dy\,\tr G_{|m_j|}(r,r;iy) \ , \\
&\tr G_{|m_j|}(r,r;iy)=\tr G_{m_j}(r,r;iy)+\tr G_{-m_j}(r,r;iy) \ .
\eal
\eeq
In the case, when there exist a set of negative discrete levels with $-1\leq \e_{n}<0$, instead of (\ref{3.18a}) one gets by analogy with Ref.~\cite{gyul1975}
\beq
\label{3.19}
\r_{VP,|m_j|}(r)= {|e|\over 2 \pi} \[\sum\limits_{m_j=\pm|m_j|}\sum\limits_{-1\leq \e_n<0}\p_{n,m_j}(r)^{\dagger}\p_{n,m_j}(r)+\frac{1}{2 \pi} \int\limits_{-\infty}^{+\infty}d y\, \tr G_{|m_j|}(r,r;iy)\].
\eeq
Proceeding further, let us mention the general properties of $\tr G_{m_j}$ under the change of the sign of external field $(Q \to -Q)$ and complex conjugation
\beq\label{3.20}
\bal
&\tr G_{-m_j}(Q;r,r;\e)=-\tr G_{m_j}(-Q;r,r;-\e) \ ,\\
&\tr G_{m_j}(Q;r,r;\e)^{\ast}=\tr G_{m_j}(Q;r,r;\e^{\ast}) \ ,
\eal
\eeq
and their direct consequence
\beq
\label{3.21}
\tr G_{m_j} (Q;r,r; i y)^{\ast}=-\tr G_{-m_j}(-Q;r,r;i y) \ .
\eeq
There follows from  (\ref{3.20}) and (\ref{3.21})
that  actually  $\r_{VP,|m_j|}(r)$  is determined by $ \mathrm{Re}\,\tr G_{|m_j|} (Q;r,r; i y)$ and so is definitely a real quantity, odd in $Q$ in accordance with the Furry theorem. In the purely perturbative region, the representation of $\r_{VP,|m_j|}(r)$ as an odd series in powers of external field (\ref{1.1}) follows directly from the Born series  for the Green function $G_{m_j}=G_{m_j}^{(0)}+G_{m_j}^{(0)} (-V) G_{m_j}^{(0)} + G_{m_j}^{(0)} (-V) G_{m_j}^{(0)} (-V) G_{m_j}^{(0)} + \dots $, whence
\beq
\label{3.211}
\mathrm{Re}\, \tr G_{m_j} (r,r; i y)=\sum\limits_{k=0} \mathrm{Re}\, \tr  \[ G_{m_j}^{(0)} \(-V G_{m_j}^{(0)}\)^{2k+1}(r,r; i y) \]  \ ,
\eeq
where $G_{m_j}^{(0)}$ is the Green function of the free radial Dirac equation with the same $m_j$.
At the same time, in presence of negative discrete levels and, moreover, in the overcritical region with $Z>Z_{cr,1}$, the oddness in $Q$ property of $\r_{VP}$ maintains \cite{gyul1975}, but now the dependence on the external field cannot be described by a power series (\ref{3.211}) any more, since  there appear in $\r_{VP}$ certain  essentially non-perturbative and so non-analytic in $Q$ components.

\section{The Divergencies and Renormalization of the Vacuum Density in 2+1 D}

The expressions (\ref{3.18})-(\ref{3.19}) for the vacuum density require renormalization, since there follows from the asymptotics of $\tr G_{m_j}$ for $r \to \inf$
\beq
\label{3.22}
\tr G_{m_j}(r,r;iy)\to\frac{i y}{\sqrt{1+y^2}}\frac{1}{r}+\frac{Q}{\(1+y^2\)^{3/2}}\frac{1}{r^2}+\mathrm{O}\(r^{-3}\), \quad r \to \inf \ ,
\eeq
 that the non-renormalized density $\r_{VP}(r)$ decreases for $r \to \inf$ as $1/r^2$, hence, the non-renormalized induced charge turns out to be logarithmically divergent.

The general result, obtained in Ref.~\cite{gyul1975} within the expansion of $\r_{VP}$ in powers of $Q$, which is valid for 1+1 and 2+1 D always, and for a spherically-symmetric external potential in the three-dimensional case, is that all the divergencies of $\r_{VP}$ originate solely from the fermionic loop with two external photon lines, which gives rise to the Uehling potential $A^{(1)}_{VP,0}$, whereas all the next-to-leading orders of expansion are already free from divergencies (see also Ref.~\cite{mohr1998} and refs. therein). In 3+1 D there diverges (logarithmically) also the fermionic loop with 4 external lines, but in the  spherically-symmetric case this divergence is not relevant for calculation of the vacuum density and vacuum energy, since within the partial expansion each term with definite momentum and parity of order $O(Q^3)$ for the density and of order $O(Q^4)$ for the energy turns out to be automatically gauge-invariant and finite without any additional regularization. Hence, in this case the renormalization procedure for the vacuum density  is actually the same for all the three spatial dimensions and reduces to the diagram with two external lines.

Thus, in order to find the renormalized density  $\r^{ren}_{VP}(r)$, the linear in $Q$ terms in the expression for $\tr G_{m_j}$ in (\ref{3.11}) should be extracted and replaced by the renormalized  perturbative density $\r^{(1)}_{VP}(r)$, determined  in (\ref{2.5c}), which corresponds to the first-order  PT and does not vanish only for $|m_j|=1/2$ (see Appendix A). For these purposes let us obtain first the component of the vacuum density $\r^{(3+)}_{VP,|m_j|}(r)$, that is defined in the next  way
\beq \label{3.23}
\bal
\r_{VP,|m_j|}^{(3+)}(r)&= {|e|\over 2 \pi} \[\sum\limits_{m_j=\pm|m_j|}\sum\limits_{-1\leq \e_n<0}\p_{n,m_j}(r)^{\dagger}\p_{n,m_j}(r)\right.\\
&\left.+\frac{1}{\pi} \int\limits_{0}^{\infty}d y\,\mathrm{Re}\( \tr G_{|m_j|}(r,r;iy)-2\, \tr G^{(1)}_{m_j}(r;i y)\)\] \ ,
\eal
\eeq
where $G^{(1)}_{m_j}(r;i y)$ is the linear in $Q$ component of the partial Green function $G_{m_j}(r,r;i y)$, which  is found from the first Born approximation for $G_{m_j}$:
\beq \label{3.231}
G^{(1)}_{m_j}=Q \( \pd G_{m_j} /\pd Q \)_{Q=0}=G^{(0)}_{m_j} (-V) G^{(0)}_{m_j} \ .
\eeq
For the external potential (\ref{1.1}) the explicit expression of $G^{(1)}_{m_j}(r;i y)$  for $r\leq R$ reads
\beq \label{3.24}
\bal
\tr G^{(1)}_{m_j}(r;i y)&=\frac{Q}{(i y-1)^2}\[\(\tilde{\g}^2K^2_{|m_j-1/2|}(\tilde{\g}r)+(1-i y)^2 K^{2}_{|m_j+1/2|}(\tilde{\g}r)\)\right.\\
&\left.\times\int\limits_{0}^{r}dr'\,\frac{r'}{R}\(\tilde{\g}^2I^2_{|m_j-1/2|}(\tilde{\g}r')+(1-i y)^2 I^{2}_{|m_j+1/2|}(\tilde{\g}r')\)\right.\\
&\left.+\(\tilde{\g}^2I^2_{|m_j-1/2|}(\tilde{\g}r)+(1-i y)^2 I^{2}_{|m_j+1/2|}(\tilde{\g}r)\) \right. \\
&\left.\times\left\lbrace\int\limits_{r}^{R}dr'\,\frac{r'}{R}\(\tilde{\g}^2K^2_{|m_j-1/2|}(\tilde{\g}r')+(1-i y)^2 K^{2}_{|m_j+1/2|}(\tilde{\g}r')\) \right. \right.\\
&\left.\left.+\int\limits_{R}^{\infty}dr'\,\(\tilde{\g}^2K^2_{|m_j-1/2|}(\tilde{\g}r')+(1-i y)^2 K^{2}_{|m_j+1/2|}(\tilde{\g}r')\)\right\rbrace\] \  ,
\eal
\eeq
while for  $r>R$
\beq \label{3.24a}
\bal
\tr G^{(1)}_{m_j}(r;i y)&=\frac{Q}{(i y-1)^2}\[\(\tilde{\g}^2K^2_{|m_j-1/2|}(\tilde{\g}r)+(1-i y)^2 K^{2}_{|m_j+1/2|}(\tilde{\g}r)\)\right.\\
&\left.\times\left\lbrace\int\limits_{0}^{R}dr'\,\frac{r'}{R}\(\tilde{\g}^2I^2_{|m_j-1/2|}(\tilde{\g}r')+(1-i y)^2 I^{2}_{|m_j+1/2|}(\tilde{\g}r')\)\right.\right.\\
&\left.\left.+\int\limits_{R}^{r}dr'\,\(\tilde{\g}^2I^2_{|m_j-1/2|}(\tilde{\g}r')+(1-i y)^2 I^{2}_{|m_j+1/2|}(\tilde{\g}r')\)\right\rbrace\right.\\
&\left.+\(\tilde{\g}^2I^2_{|m_j-1/2|}(\tilde{\g}r)+(1-i y)^2 I^{2}_{|m_j+1/2|}(\tilde{\g}r)\)\right.\\
&\left.\times\int\limits_{r}^{\infty}dr'\,\(\tilde{\g}^2K^2_{|m_j-1/2|}(\tilde{\g}r')+(1-i y)^2 K^{2}_{|m_j+1/2|}(\tilde{\g}r')\)\] \ ,
\eal
\eeq
where $\tilde{\g}=\sqrt{1+y^2}$. The integrals, entering into the expressions (\ref{3.24}) and (\ref{3.24a}) for $\tr G^{(1)}_{m_j}$,  could also  be in principle evaluated  analytically, but  they are not given here explicitly because of their cumbersome form.

So the renormalized vacuum density takes the form
\beq\label{3.25}
\r^{ren}_{VP}(r)=2\[\r_{VP}^{(1)}(r)+\sum\limits_{m_j=1/2,\,3/2,..}\r_{VP,|m_j|}^{(3+)}(r)\] \ ,
\eeq
where  $\r_{VP}^{(1)}(r)$ is the perturbative vacuum density (\ref{2.2}), evaluated by means of the polarization function (\ref{2.4}) in the first order of PT. Such expression for $\r^{ren}_{VP}(r)$ guarantees the vanishing  total vacuum charge $Q^{ren}_{VP}=\int\limits d^{2}r\,\r^{ren}_{VP}(r)$ for $Z<Z_{cr,1}$, since  $Q_{VP}^{(1)}$ is zero by construction, while the subsequent direct check confirms that the contribution of $\r^{(3+)}_{VP,|m_j|}(r)$ to  $Q^{ren}_{VP}$ for $Z<Z_{cr,1}$ vanishes too. Unlike 1+1 D,  in 2+1 D such a check cannot  be performed in the purely analytical form any more  due to complexity of  expressions, entering into $\r^{(3+)}_{VP,|m_j|}(r)$. Nevertheless, it could be quite reliably performed via combination of analytical and numerical methods (see Appendix B).  Moreover, it suffices to verify the disappearance of the total charge $Q_{VP}^{ren}$ not for the entire subcritical region, but only in absence of negative discrete levels. In presence of the latter,  the vanishing  total charge for $Z<Z_{cr,1}$ follows from model-independent arguments, which are based on the starting expression for the vacuum density (\ref{3.1}). There follows from (\ref{3.1}) that the change of the integral induced charge is possible for  $Z>Z_{cr,1}$ only, when the  discrete levels reach the lower continuum. Moreover, each doubly degenerate energy level, diving into the lower continuum,  yields the change of the integral charge exactly by  $(-2|e|)$.  One of the possible correct ways to prove this statement is given in Ref.~\cite{davydov2017}.  Let us specially mention that this effect is essentially non-perturbative and  completely included in $\r_{VP}^{(3+)}$, while  $\r_{VP}^{(1)}$ does not participate in it and still makes a purely vanishing contribution into the total induced charge. Thus, the behavior of the renormalized by means of (\ref{3.25}) vacuum density in the non-perturbative region turns out to be exactly such that should be expected from the  general assumptions about the structure of the electron-positron vacuum for $Z> Z_{cr}$ \cite{greiner1985,plunien1986,greiner2012,raf2017}.

A more detailed picture of changes, taking place in  $\r^{ren}_{VP}(r)$, turns out to be quite similar to that considered in Refs.~\cite{raf2017}, \cite{greiner1985}\nocite{plunien1986}-\cite{greiner2012} for 3+1 D, based on the U.~Fano approach \cite{fano1962}. The main result is that, when the discrete level  $\p_{n,m_j}(r)$ reaches the lower continuum, the change in the vacuum density equals to
\beq\label{3.26}
\D\r_{VP}(r)=\left. -2\, |e|\, \p_{n,m_j}(r)^{\dagger}\p_{n,m_j}(r)\right|_{\e_n=-1} \ .
\eeq
However, it should be noted that this approach contains a number of approximations too, and so the expression (\ref{3.26}) is exact only in the  vicinity of the corresponding $Z_{cr}$, what is quite explicitly shown in Refs.~\cite{davydov2017}\nocite{sveshnikov2017}-\cite{voronina2017}.

The substantial difference between vacuum polarization  effects in 2+1 and 3+1 D and those from the one-dimensional case shows up also in that the renormalized density (\ref{3.25}) is represented now as a infinite partial series in $m_j$. Hence, there appears the natural question concerning the convergence of this series. To answer this question let us consider the asymptotics of $\r_{VP,|m_j|}^{(3+)}(r)$  for large $|m_j|$. In the first step, the integrand in (\ref{3.23}) should be explored. The asymptotics of $\tr G_{m_j}(r,r;iy)$  for $|m_j|\to \infty$ takes the form
\beq  \label{3.26a}
\tr G_{m_j}(r,r;iy)\to \left\lbrace \bal
&\frac{iy+V_0}{|m_j|}+\frac{\mathrm{sgn}(m_j)}{m_j^2}+\mathrm{O}\(|m_j|^{-3}\), \quad &r\leq R \ , \\
&\frac{iy+Q/r}{|m_j|}+\frac{\mathrm{sgn}(m_j)}{m_j^2}+\mathrm{O}\(|m_j|^{-3}\), \quad &r> R \ .
\eal \right.
\eeq
At the same time, the asymptotics of $\tr G^{(1)}_{m_j}(r;iy)$ (\ref{3.24}), (\ref{3.24a}) for large $|m_j|$ is the following
\beq \label{3.26b}
\tr G^{(1)}_{m_j}(r;iy)\to \left\lbrace \bal
&V_0\frac{1}{|m_j|}+\mathrm{O}\(|m_j|^{-3}\), \quad &r\leq R \ , \\
&\frac{Q}{r}\frac{1}{|m_j|}+\mathrm{O}\(|m_j|^{-3}\), \quad &r> R \ .
\eal \right.
\eeq
From (\ref{3.26a}) and (\ref{3.26b}) with account of definition of $\tr G_{|m_j|}(r,r;iy)$ (\ref{3.18a}) one obtains:
\beq \label{3.26c}
\mathrm{Re}\[ \tr G_{|m_j|}(r,r;iy)-2\, \tr G^{(1)}_{m_j}(r;i y)\]=\mathrm{O}\(|m_j|^{-3}\), \quad |m_j|\to \infty.
\eeq
Moreover, with increasing $|m_j|$ the  discrete levels necessarily rise, so for the given $Q$ and $|m_j|\to\infty$ the negative discrete levels should be absent at all.  Hence, in this limit the sum $\sum \p_{\e,m_j}(r)^{\dagger}\p_{\e,m_j}(r)$ in the expression (\ref{3.23}) disappears. Since the integral over $dy$ in (\ref{3.23}) converges uniformly with respect to  $|m_j|$ and $r$, considered as the external parameters (see Appendix C), there follows from (\ref{3.26c}) that $\r_{VP,|m_j|}^{(3+)}(r)$ for $|m_j| \to \infty$ behaves like $\mathrm{O}(|m_j|^{-3})$. So the partial series in (\ref{3.25})  converges and the renormalized vacuum density $\r^{ren}_{VP}(r)$ turns out to be finite everywhere besides the logarithmic singularity at $r=R$, originating from $\r_{VP}^{(1)}(r)$.

In the Fig. 2 for the external potential (\ref{1.00}) with the parameters $\a$ and $R=R(Z)$, specified above,  the renormalized vacuum density $\r_{VP}^{ren}(r)$ is shown for the purely perturbative region with $Z=10$, thereafter for  $Z=108$, when the first $Z_{cr,1}\simeq 108.1$ is not reached yet, then  for  $Z=109$, when the first discrete level has just dived into the lower continuum, further for $Z=133$, when the second critical $Z_{cr,2}\simeq 133.2$ is not reached yet, and finally for $Z=134$, i.e. just after the second discrete level diving into  the lower continuum. The critical charges are found from the transcendental equation, which follows from matching of regular solutions for $r<R$ (\ref{3.8}) and $r>R$ (\ref{3.9}) at $\e=-1$:
\beq\label{3.261}
\bal
 &I_{|m_j-1/2|}\(\x R\)\[K_{2s-1}\(\sqrt{8 Q R}\)+K_{2s+1}\(\sqrt{8 Q R}\)+\right.\\
&\left.+ \sqrt{\frac{2}{Q R}}m_j K_{2s}\(\sqrt{8 Q R}\)\]+\sqrt{2(2-V_0)}\,I_{|m_j+1/2|}\(\x R\)K_{2s}\(\sqrt{8 Q R}\)=0 \ .
 \eal
 \eeq
In (\ref{3.261}), $K_{\n}(z)$ is the  Macdonald function, into which the Whittaker functions transform for $\e \to -1$. The  numerical integration confirms that the total vacuum charge for $Z=10, \ 108$ equals to zero, for $Z=109, \ 133$ due to above-mentioned twofold degeneracy it equals to $(-2 |e|)$, while for $Z=134$ --- to $(-4 |e|)$  correspondingly (for details of calculations see Appendix B). At $r=R$ $\r_{VP}^{ren}(r)$ reveals a logarithmic singularity, which is caused by the contribution of $ \r_{VP}^{(1)}(r)$  (Fig. 3). At the same time,  $ \r_{VP}^{(3+)}(r)=\sum_{m_j}\r_{VP,|m_j|}^{(3+)}(r) $  turns out to be a continuous function everywhere, what is shown in Fig. 4. Because  it is difficult to see from the Fig. 4 that $\int \! r dr \ \r_{VP}^{(3+)}(r)$ for $Z=10$ and $Z=108$ vanishes, in the Fig. 4a we demonstrate  that $ \r_{VP}^{(3+)}(r)$ is actually a sign-alternating function of $r$ for these $Z$. Let us also mention that in the overcritical region with increasing  $Z$ the change of $\r_{VP}^{ren}(r)$  proceeds not only in a discrete manner due to vacuum shells formation from the discrete levels diving into the lower continuum, but also continuously, due to the changes in the density of states of the continuous spectrum and evolution of discrete levels with growing $Z$.

\begin{center}
	\includegraphics[scale=0.52]{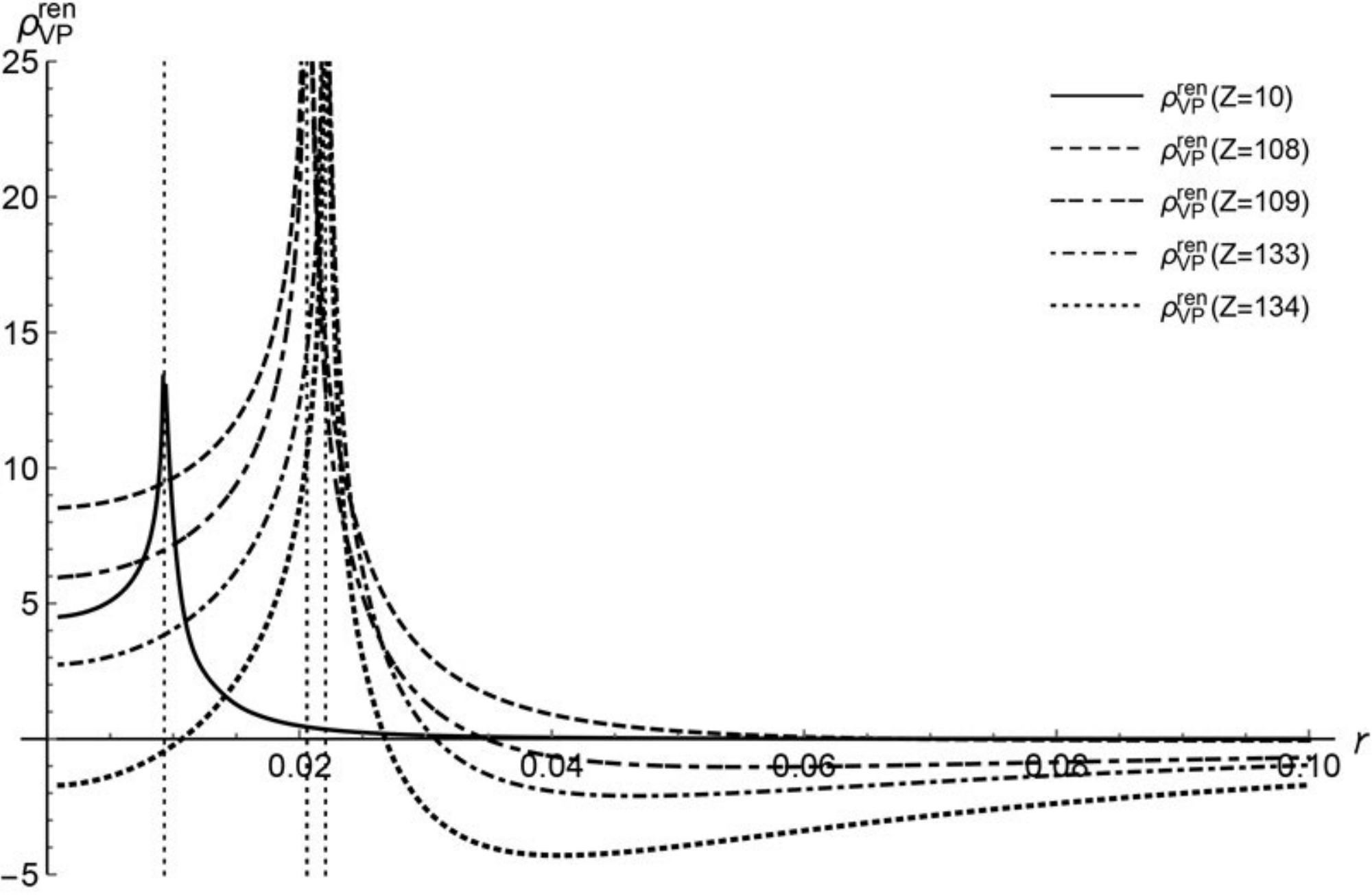} \\
	{\small Fig.2. $\r_{VP}^{ren}(r)$ for $Z=10, \ 108, \ 109, \ 133, \ 134 $.}
\end{center}

\begin{center}
	\includegraphics[scale=0.52]{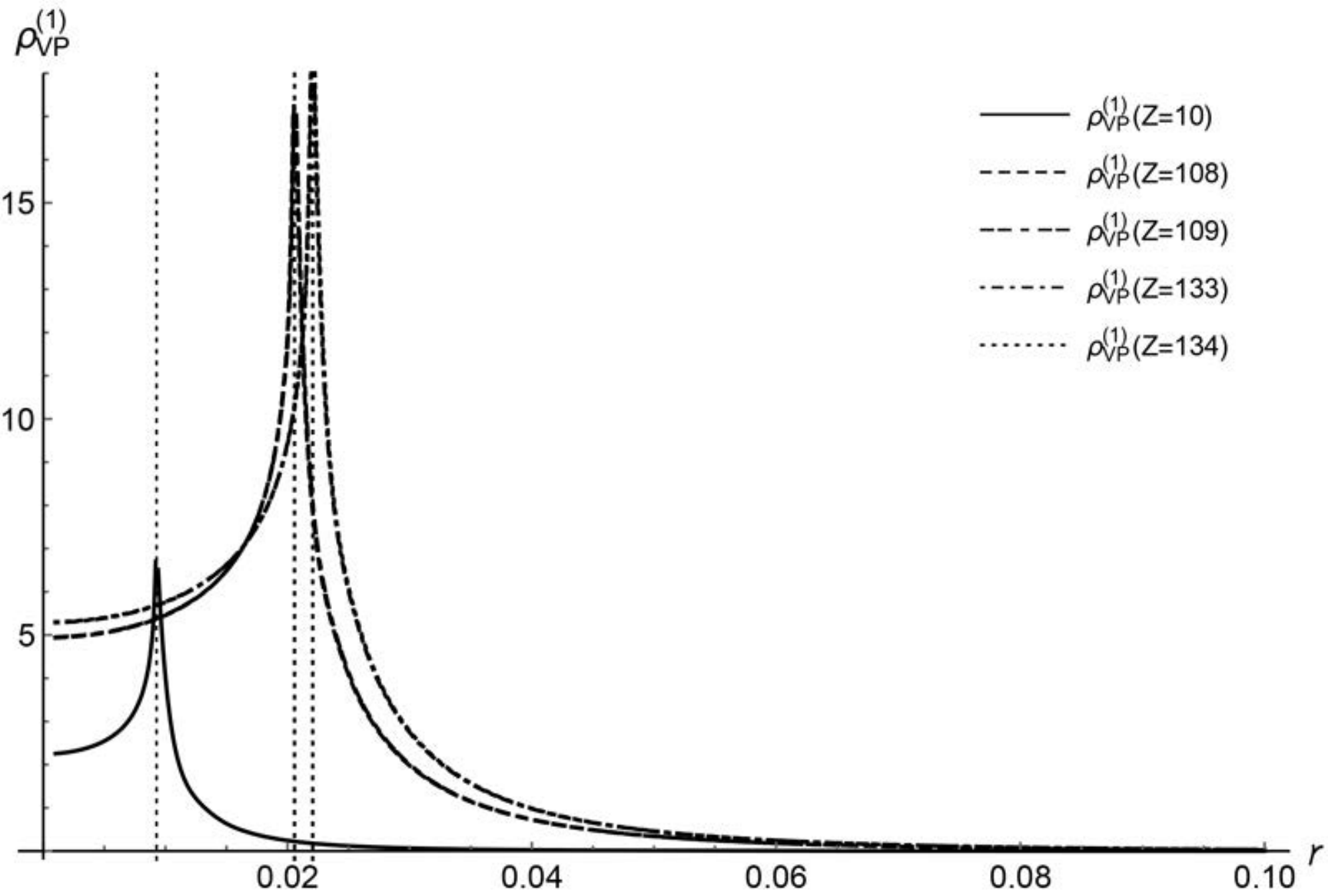} \\
	{\small Fig.3. $\r_{VP}^{(1)}(r)$ for $Z=10, \ 108, \ 109, \ 133, \ 134 $.}
\end{center}

\begin{center}
	\includegraphics[scale=0.52]{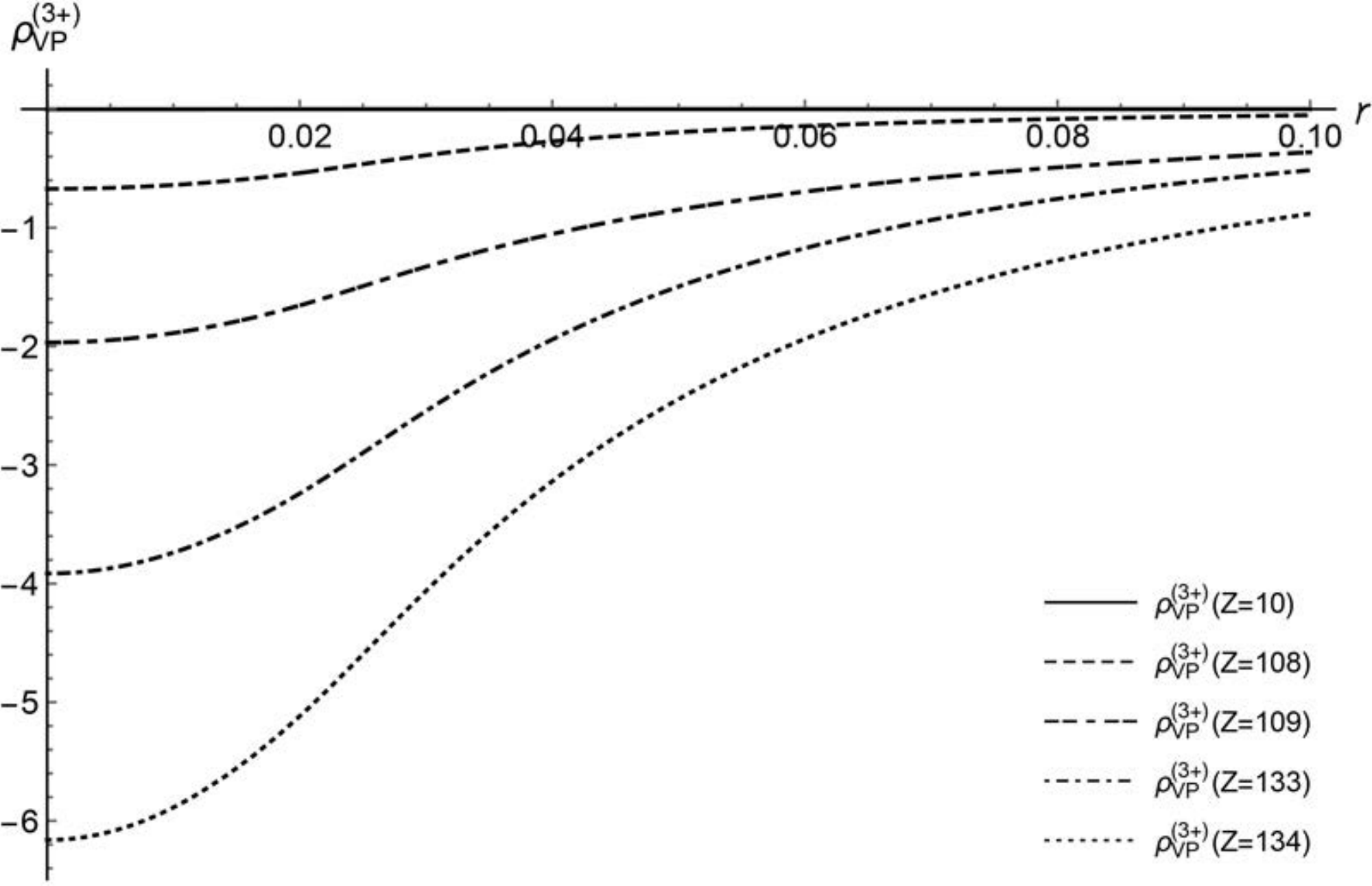} \\
	{\small Fig.4. $\r_{VP}^{(3+)}(r)$ for $Z=10, \ 108, \ 109, \ 133, \ 134 $.}
\end{center}

\begin{center}
	\includegraphics[scale=0.52]{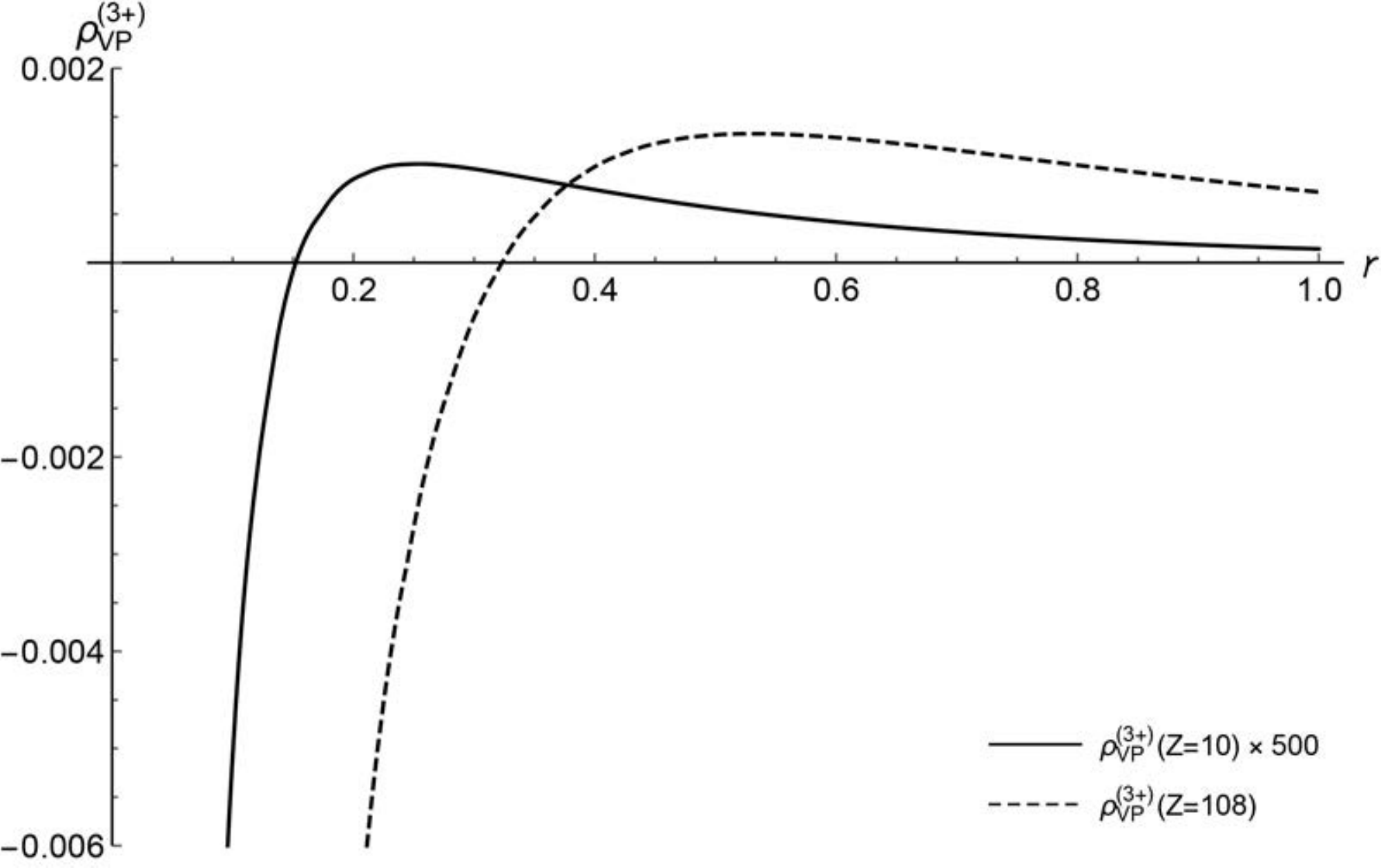} \\
	{\small Fig.4a. $\r_{VP}^{(3+)}(r)$ for $Z=10$ and $Z=108$ as sign-alternating functions of $r$.}
\end{center}
It should be also noted that for undercritical $Z$ the main contribution to the partial expansion (\ref{3.25}) for  $\r_{VP}^{ren}(r)$ is given by the term $\r_{VP,|m_j|=1/2}^{(3+)}$ (see Fig. 5). At the same time, for overcritical $Z$ the sum (\ref{3.25}) depends first of all on the contribution from those  $|m_j|$, for which at least one level has already sunk into the lower continuum.
\begin{center}
	\includegraphics[scale=0.5]{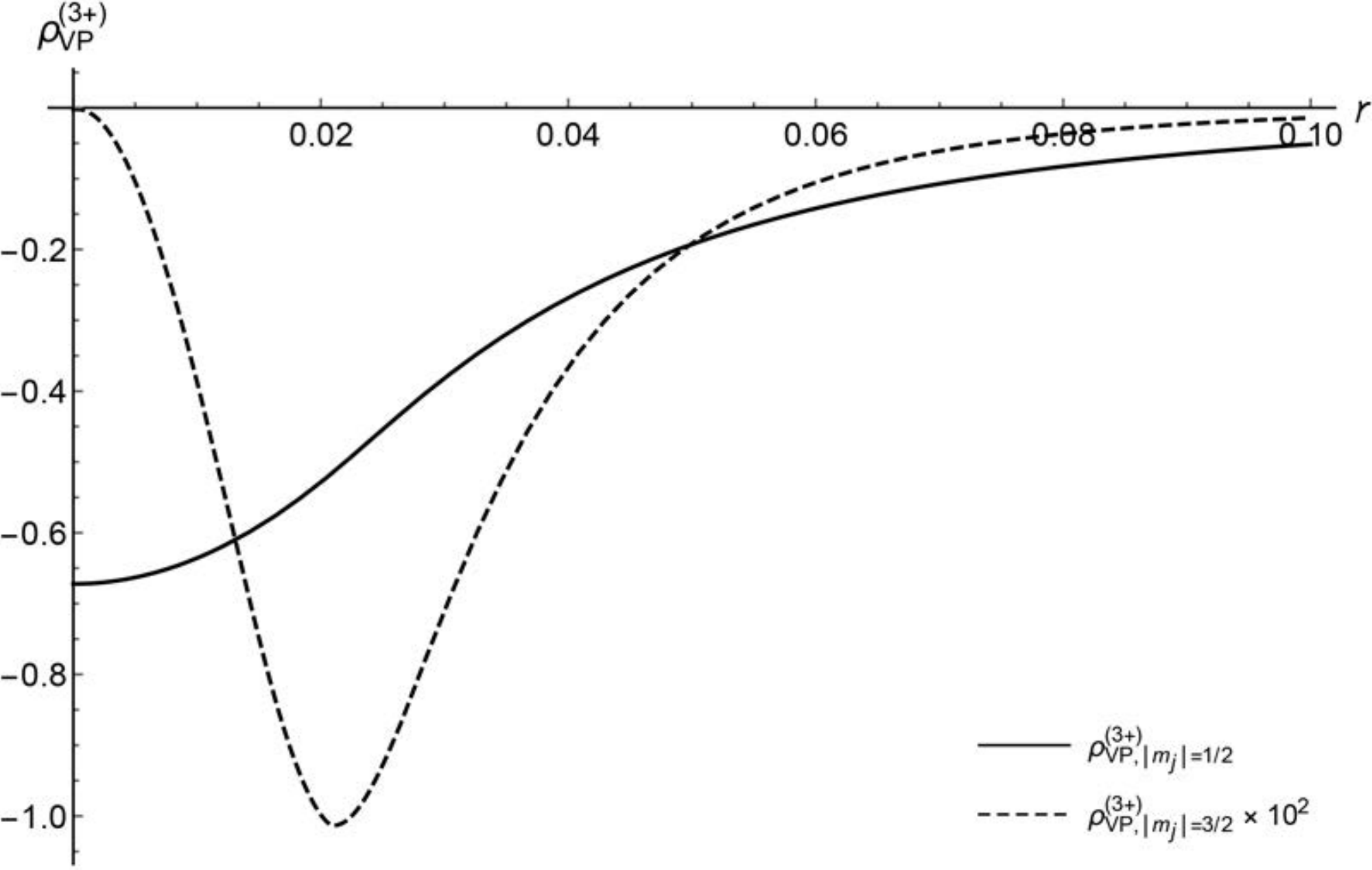} \\
	{\small Fig.5. $\r_{VP,|m_j|}^{(3+)}(r)$ at $Z=108$ for $|m_j|=1/2$ and $|m_j|=3/2$.}
\end{center}

 Thus, the correct approach to evaluation of  $\r_{VP}^{ren}(r)$  for all regions of $Z$ consists in making use of relations (\ref{3.23}) and (\ref{3.25}) with subsequent check of  the expected integer value of the induced charge via direct numerical integration of $\r_{VP}^{ren}(r)$.
The appearing general picture for the renormalized vacuum density $\r_{VP}^{ren}(r)$ and its components in the range $0< Z <1000$ is shown in the Figs.6-8  for  the four most representative values $ Z=100\, , 200\, , 500\, , 1000 $. The almost vertical pics on the curves $\r_{VP}^{ren}(r)$ in the Fig.6 correspond to the logarithmic singularities in the vacuum density, caused by $\r_{VP}^{(1)}(r)$.
\begin{center}
	\includegraphics[scale=0.52]{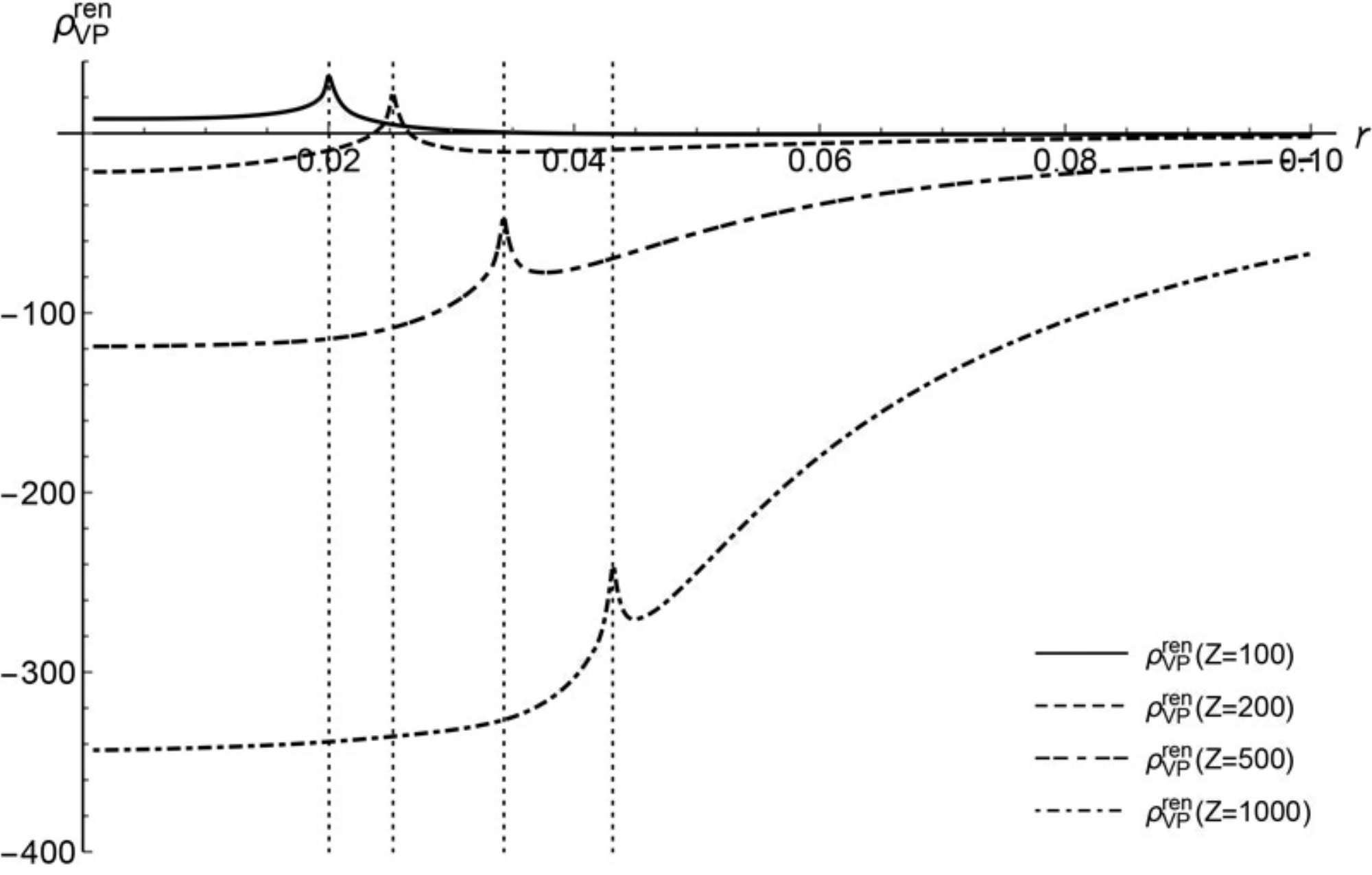} \\
	{\small Fig.6. $\r_{VP}^{ren}(r)$ for $Z=100, \ 200, \ 500, \ 1000 $.}
\end{center}

\begin{center}
	\includegraphics[scale=0.52]{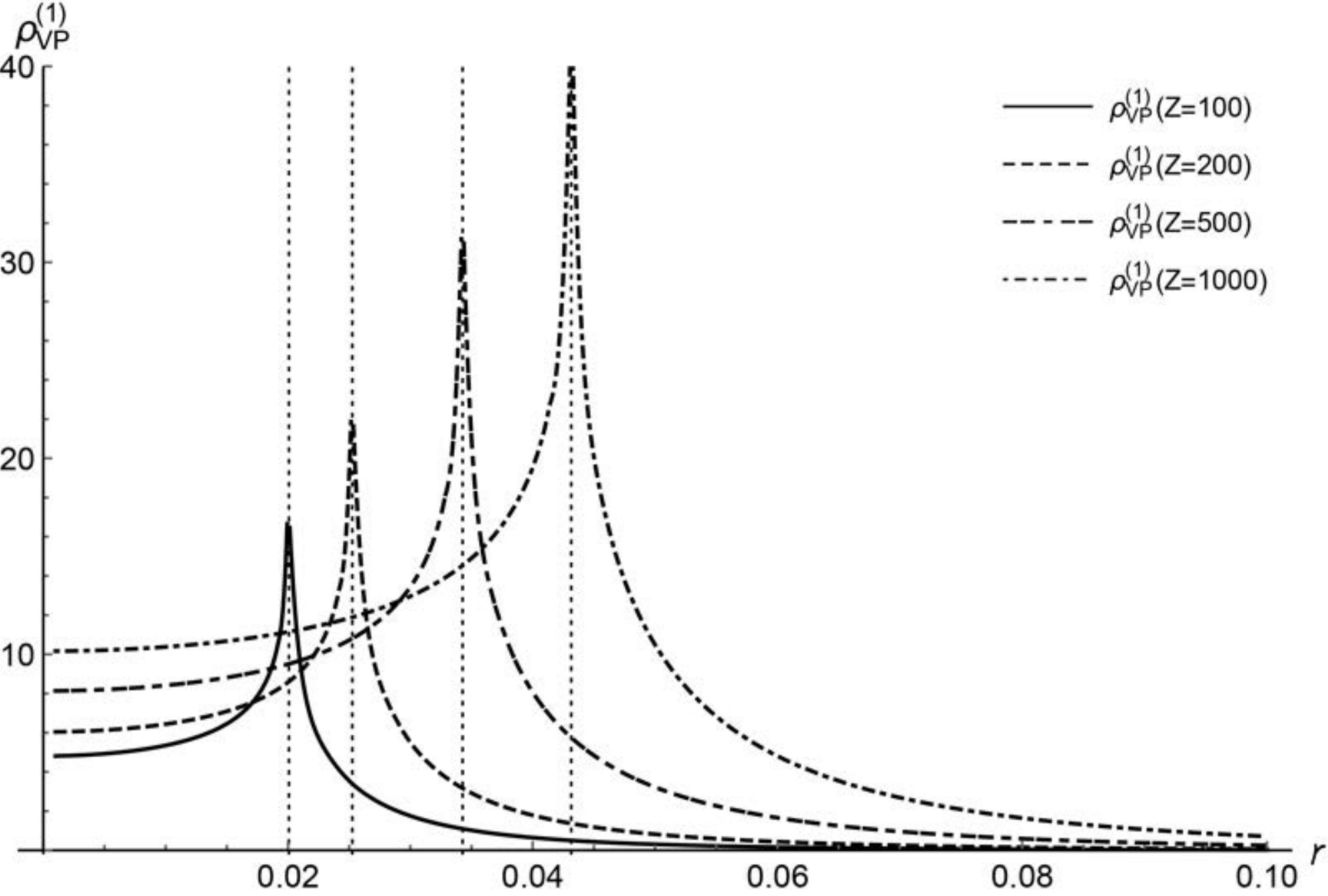} \\
	{\small Fig.7. $\r_{VP}^{(1)}(r)$ for $Z=100, \ 200, \ 500, \ 1000 $.}
\end{center}

\begin{center}
	\includegraphics[scale=0.52]{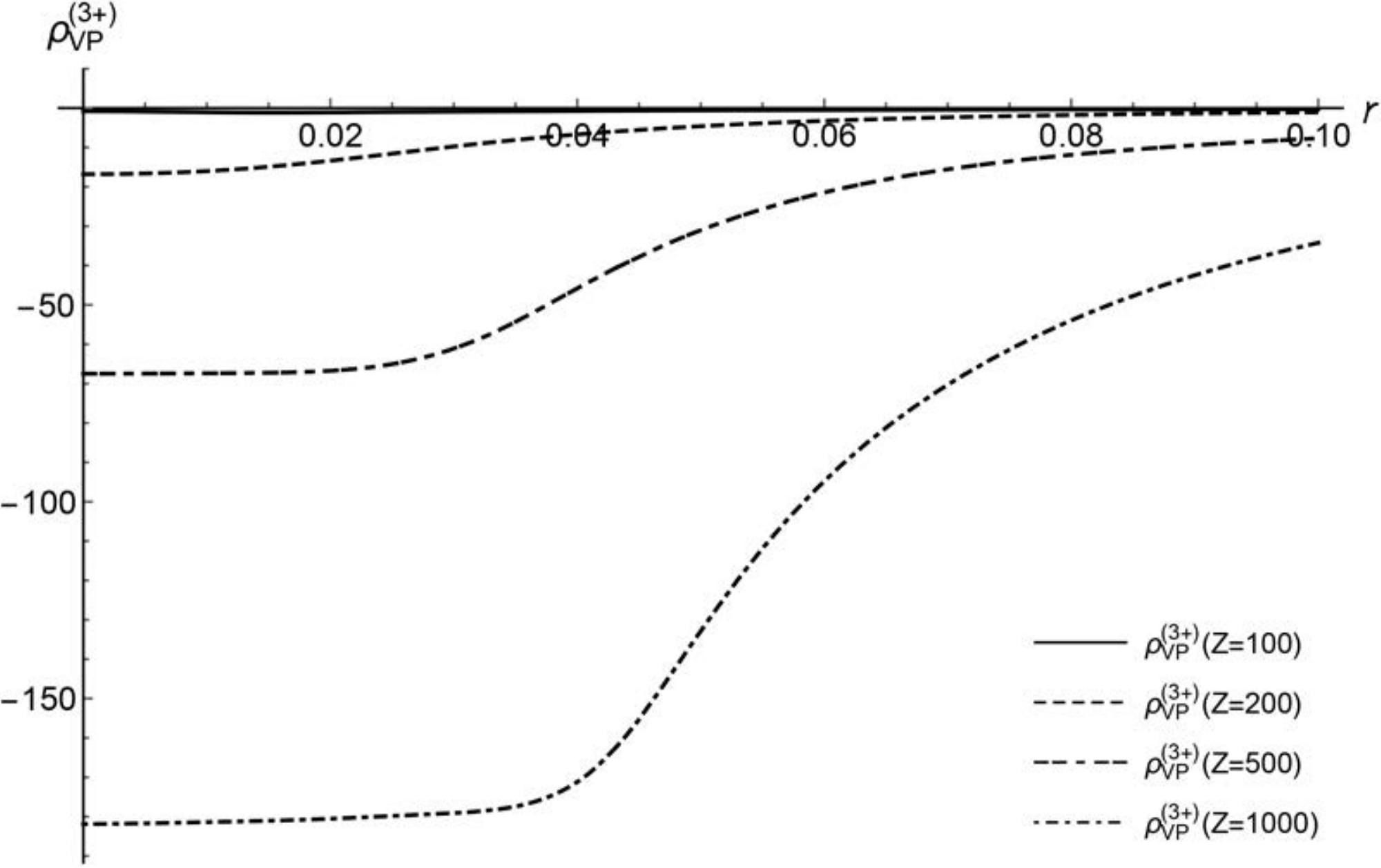} \\
	{\small Fig.8. $\r_{VP}^{(3+)}(r)$ for $Z=100, \ 200, \ 500, \ 1000 $.}
\end{center}
Moreover, it should be clear that the main contribution to the vacuum density for the given  $Z$ is produced by those partial channels, in which the discrete levels have already reached the lower continuum. The histograms, demonstrating the contribution from different partial channels to the total induced vacuum charge $Q_{VP}^{ren}(Z) $, are shown in the Fig.9 for $Z=500$ and $Z=1000$.
\begin{figure}[h!]
\begin{minipage}{0.48\linewidth}	
\center{\includegraphics[scale=0.7]{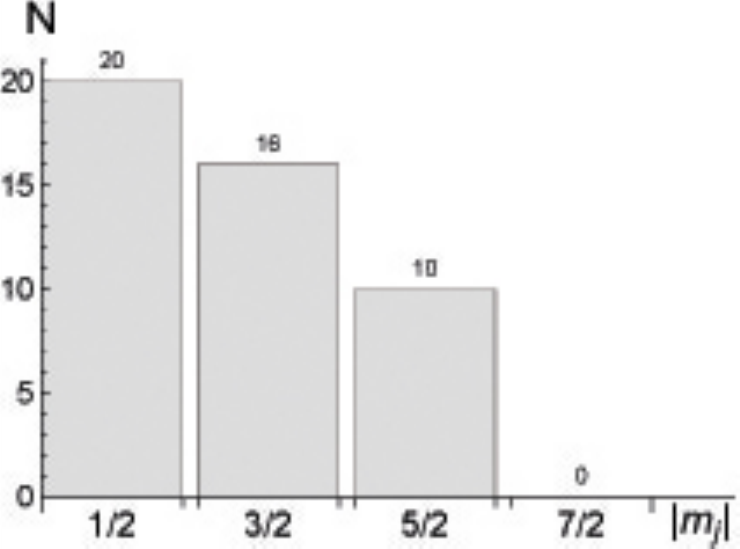} \\ a) }
\end{minipage}
\hfill
\begin{minipage}{0.48\linewidth}
\center{\includegraphics[scale=0.76]{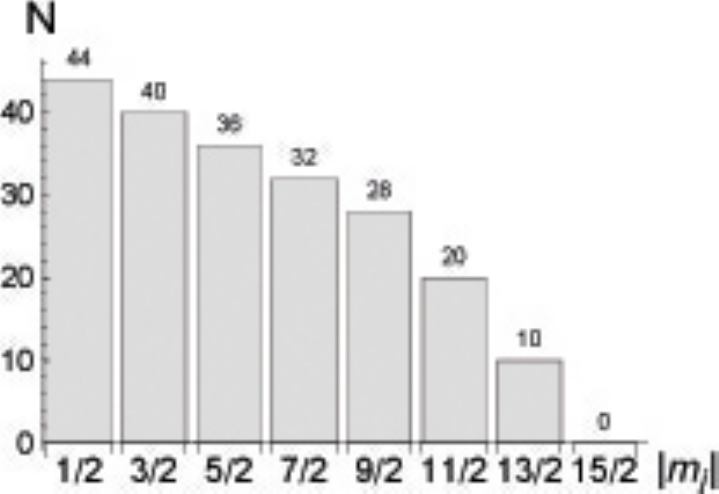} \\ b) }
\end{minipage}
\begin{center}
		{\small Fig.9. The contribution from different $ m_j $ to the vacuum polarization for (a):$Z=500$; (b):$Z=1000$ \ .}
	\end{center}
\end{figure}
\begin{figure}[h!]
\begin{minipage}{0.48\linewidth}	
\center{\includegraphics[scale=0.65]{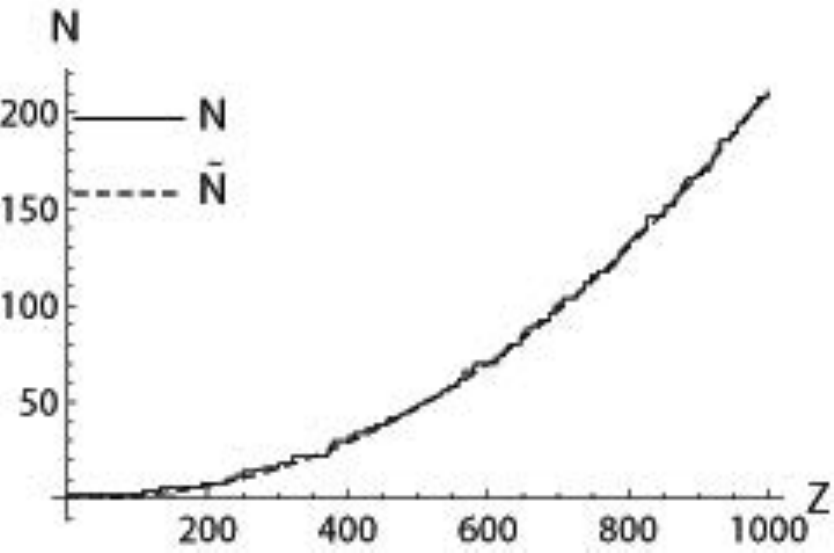} \\ a) }
\end{minipage}
\hfill
\begin{minipage}{0.48\linewidth}
\center{\includegraphics[scale=0.65]{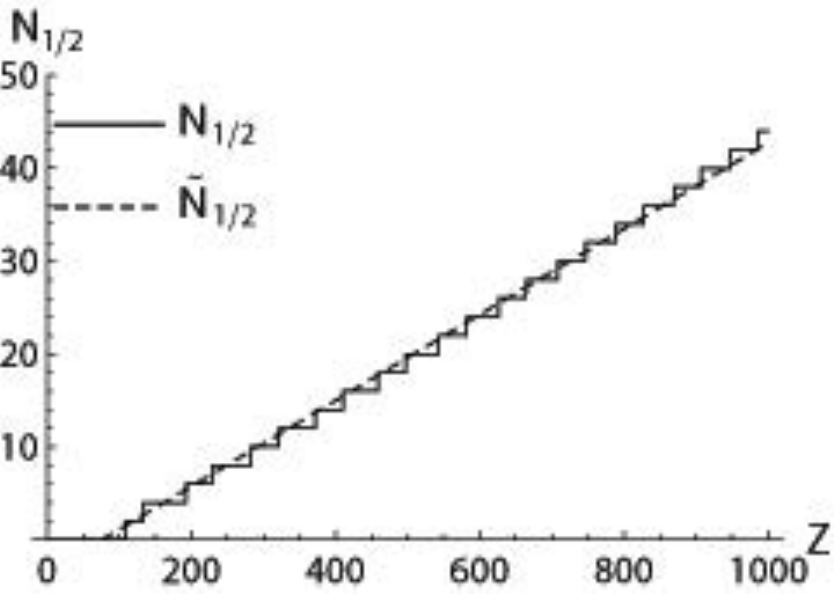} \\ b) }
\end{minipage}
\begin{center}
		{\small Fig.10. (a): The total number of vacuum shells $N(Z)$ and its approximation via power function ${\tilde N}(Z)$; (b): the number of vacuum shells $N_{1/2}(Z)$ and its approximation via power function ${\tilde N}_{1/2}(Z)$\ .}
	\end{center}
\end{figure}	
In the Fig.10a  both the dependence on $Z$ for the total number $N(Z)$ of vacuum shells (from all  $m_j$), formed from levels, which have already sunk into the lower continuum, and its power-like approximation ${\tilde N}(Z)=6.69 \times 10^{-5} \times Z^{2.17}$, are shown. As it follows from the Fig.10a, in 2+1 D the growth rate of the shells total number for $Z \gg Z_{cr,1}$  becomes already more than quadratic. Moreover, the  power $Z^{2.17}$  estimates the growth rate of $N(Z)$ from below, since only the range  $0 < Z < 1000$ is used for this approximation. By extending the approximation range in  $Z$ to the right the growth rate increases, but such  $Z$ in this work are not considered by default. In any case, however, the growth rate of shells number for $Z \gg Z_{cr,1}$   more than  quadratic one significantly influences on the behavior of the vacuum energy in the overcritical region in 2+1 D, since now the shell effect becomes the dominant factor in $\E_{VP}^{ren}(Z)$ for $Z \gg Z_{cr,1}$. In the Fig.10b the dependence of the shells number in the lowest channel $N_{1/2}(Z)$ on $Z$ due to levels with $|m_j|=1/2$, which have sunk into the lower continuum, and its power-like approximation  ${\tilde N}_{1/2}(Z)=-3.53+0.046 Z \ $, are presented. The last result shows that in its structure the partial channel with $|m_j|=1/2$ is practically identical with $\r_{VP}^{ren}(Z)$ for the one-dimensional case, while the growth rate of the vacuum shells number in this channel, as in 1+1 D, is close to the linear one, at least in the considered range of external parameters.

\section{Conclusion}

Thus, in the present part I of our work the  vacuum polarization  effects in the  2+1-dimensional QED for a model potential  (\ref{1.00}) have been considered at the level of the renormalized vacuum density $\r_{VP}^{ren}(Z)$. The 2+1-dimensional case differs significantly from the one-dimensional one first of all in that  $\r_{VP}^{ren}(Z)$ is represented now by an infinite  series in the rotational quantum number $m_j$, and so there appears an additional problem of its convergence.  As it was shown in this paper, this problem could be also solved successfully by renormalization of the vacuum density within PT, i.e. via regularization of the solely divergent Feynman graph in the form of the fermionic loop with two external lines. Simultaneously, the integral vacuum charge vanishes automatically in the subcritical region, and is changed  by $(-2|e|)$  upon diving of each subsequent doubly degenerate  discrete level into the lower continuum. Such behavior of  $\r_{VP}^{ren}(Z)$ in the overcritical region confirms once more the assumption of the neutral vacuum transmutation  into the charged one under such conditions \cite{raf2017,rg1977,greiner1985, plunien1986,greiner2012}, and thereby of the spontaneous positron emission,  accompanying the emergence of the next vacuum shell due to the total charge conservation.

It would be also worth to note that in its structure the partial channel with $|m_j|=1/2$ is almost identical with $\r_{VP}^{ren}(Z)$ for the one-dimensional case, and the growth rate of the vacuum shells number in this channel, as in 1+1 D,  does not exceed $\sim Z^s \ , \ 1<s<2$, at least in the considered range of the external parameters (see Fig.10b). However, due to the contribution from the whole partial series for $\r_{VP}^{ren}(Z)$, in the overcritical region the growth rate of the total vacuum shells number   turns out to be sufficiently higher (see Fig.10a). The estimate given above is not less than $\sim Z^{2.17}$, since for $Z \gg Z_{cr,1}$ the levels with $|m_j|>1/2$ reach the lower continuum too. And in the following part II of our work we will show that the origin of the essentially nonlinear behavior of the vacuum energy $\E_{VP}^{ren}(Z)$ for $Z>Z_{cr,1}$ is indeed  the non-perturbative changes in the vacuum charge density due to the increasing with $Z$ total number of discrete levels from different partial channels, diving into the lower continuum (``the shell effect''), and so in 2+1 D the rate of decrease of  $\E_{VP}^{ren}(Z)$ turns out to be  sufficiently larger than $\sim -\eta\, Z^2$.

\appendix
\numberwithin{equation}{section}

\section{The Uehling Potential in 2+1 D}

Here we consider in detail the derivation of the Uehling potential in 2+1 D for a static external field $A_0^{ext}(r)$ in the form (\ref{1.00}). The starting point is the expression (\ref{2.3})
\beq
A_{VP,0}^{(1)}(\vec{r})=\int\!\! \frac{d^2 q}{(2\pi)^2} \mathrm{e}^{i \vec{q} \vec{r}} \Pi_{R}\(-q^2\)\tilde{A}^{ext}_{0}\(\vec{q}\),\qquad \tilde{A}_{0}\(\vec{q}\)=\int\!\! d^2y\, \mathrm{e}^{-i\vec{q}\vec{y}}A_{0}^{ext}(y) \ .
\eeq
Proceeding further, we recall the plane wave expansion in cylindrical functions
\beq
\mathrm{e}^{i k r \cos \vf} = \sum_{m_l=-\infty}^{+\infty} i^{m_l} J_{m_l}(k r) \mathrm{e}^{i m_l \vf} \ ,
\eeq
whence
\beq
\bal
\tilde{A}_{0}\(\vec{q}\)&=\int\limits_{0}^{2\pi}d\vf_y \int\limits_{0}^{\infty}dy \,y\, \mathrm{e}^{-i q y \cos \vf_y}A_{0}^{ext}(y)\\
&=\sum_{m_l=-\infty}^{+\infty} (-i)^{m_l} \int\limits_{0}^{\infty}dy \,y\,J_{m_l}(q y) A_{0}^{ext}(y) \int\limits_{0}^{2\pi}d\vf_y\,\mathrm{e}^{i m_l \vf_y} \\
&=\sum_{m_l=-\infty}^{+\infty} (-i)^{m_l} 2 \pi \d_{m_l,0}\int\limits_{0}^{\infty}dy \,y\,J_{m_l}(q y) A_{0}^{ext}(y)=2\pi\,\int\limits_{0}^{\infty}dy \,y\,J_{0}(q y) A_{0}^{ext}(y)\\
&=2 \pi Z |e|\[\frac{1}{R}\int\limits_{0}^{R}dy \,y\,J_{0}(q y)+\int\limits_{R}^{\infty}dy \,J_{0}(q y)\]\\
&=\frac{\pi Z|e|}{q}\(2\[1+J_1(q R)-qR J_0(q R)\]+\pi q R\[ J_0(q R) \mathbf{H}_1(q R)- J_1(q R) \mathbf{H}_0(q R)\]\) \ ,
\eal
\eeq
where $\mathbf{H}_\n(z)$ is the Struve function. As a result,
\beq
\bal
A_{VP,0}^{(1)}(\vec{r})&=\frac{Z |e| \a }{8 \pi}\int\limits_{0}^{\infty}dq\,\frac{1}{q}\,\int\limits_{0}^{2\pi}d\vf_q\, \mathrm{e}^{i q r \cos \vf_q}\[\frac{2 }{q} + \(1-\frac{4}{q^2}\) \arctg \( \frac{q}{2}\)\]\\
&\times \(2\[1+J_1(q R)-q R J_0(q R)\]+\pi q R\[ J_0(q R) \mathbf{H}_1(q R)- J_1(q R) \mathbf{H}_0(q R)\]\)\\
&=\frac{Z |e| \a }{8 \pi}\sum_{m_l=-\infty}^{+\infty}i^{m_l}\int\limits_{0}^{\infty}dq\,\frac{J_{m_l}(q r)}{q}\,\[\frac{2 }{q} + \(1-\frac{4}{q^2}\) \arctg \( \frac{q}{2}\)\] \\
&\times\(2\[1+J_1(q R)-q R J_0(q R)\]+\pi q R\[ J_0(q R) \mathbf{H}_1(q R)- J_1(q R) \pmb{H}_0(q R)\]\)\\
&\times\int\limits_{0}^{2\pi}d\vf_q\, \mathrm{e}^{i m_l \vf_q}\\
&=\frac{Z |e| \a }{8 \pi}\sum_{m_l=-\infty}^{+\infty}2\pi \d_{m_l,0}\int\limits_{0}^{\infty}dq\,\frac{J_{m_l}(q r)}{q}\,\[\frac{2 }{q} + \(1-\frac{4}{q^2}\) \arctg \( \frac{q}{2}\)\]\\
&\times \(2\[1+J_1(q R)-q R J_0(q R)\]+ \pi q R\[ J_0(q R) \mathbf{H}_1(q R)- J_1(q R) \mathbf{H}_0(q R)\]\)\\
&=\frac{Z |e| \a }{4}\int\limits_{0}^{\infty}dq\,\frac{J_{0}(q r)}{q}\,\[\frac{2 }{q} + \(1-\frac{4}{q^2}\) \arctg \( \frac{q}{2}\)\]\\
&\times \(2\[1+J_1(q R)-q R J_0(q R)\]+\pi q R\[ J_0(q R) \mathbf{H}_1(q R)- J_1(q R) \mathbf{H}_0(q R)\]\) \ ,
\eal
\eeq
that is the final answer for the Uehling potential (\ref{2.5}) in the case of the external field (\ref{1.00}). It should be specially mentioned that there are only the terms with $m_l=0$ in the expansion of $A_{VP,0}^{(1)}(r)$ (hence, of both $\r_{VP}(r)$ and $\E_{VP}$) in cylindrical waves, what should be quite clear, since $A_{VP,0}^{(1)}(r)$ is assembled exclusively from two axially symmetric functions --- the external field (\ref{1.00}) and the polarization function (\ref{2.4}). So within the partial expansion for $\r_{VP}$ and $\E_{VP}$ in $m_j=m_l+m_s$ the perturbative contributions $\r_{VP}^{(1)}$ and $\E_{VP}^{(1)}$  correspond always only to $|m_j|=1/2$.

\section{Explicit Calculation of  $Q^{ren}_{VP}$}

Here we will show  that in the  problem under consideration with the external field (\ref{1.00}) the renormalized induced charge $Q^{ren}_{VP}$ vanishes in the subcritical region $Z<Z_{cr,1}$. The initial expression for  $Q^{ren}_{VP}$ reads
\beq \label{q1}
Q^{ren}_{VP}=\int\limits d^{2}r\,\r^{ren}_{VP}(r)=Q^{(1)}_{VP}+Q^{(3+)}_{VP} \ ,
\eeq
where
\beq \label{q2}
Q^{(1)}_{VP}=2\int\limits d^2 r\, \r^{(1)}_{VP}(r), \quad Q^{(3+)}_{VP}=2 \int\limits d^2 r \sum_{m_{j}=1/2,\,3/2,..}\r^{(3+)}_{VP, |m_{j}|}(r) \ .
\eeq
First of all, it should be noted that vanishing of $Q^{(1)}_{VP}$ in the problems of such type with decreasing in the spatial infinity Coulomb field follows from sufficiently more general considerations (under the same conditions of absence of any special boundary conditions and/or nontrivial topology of the field manifold), and so does not require for direct integration of the expression (\ref{2.5c}).

 For simplicity let us consider the 1+1 D problem, although such analysis could be easily performed in 2+1 and 3+1 D in presence of a static external field too. In the momentum space (up to factors like $2 \pi$ and general signs) the static equation for the potential $A_0(q)$, generated by the charge density $\r(q)$, takes the form
\beq \label{q3}
(q^2-{\tilde \Pi_R}(q))A_0(q)=\r(q) \ ,
\eeq
where ${\tilde \Pi_R}(q)$ is the polarization operator in the standard form with dimension $[q^2]$, which is introduced via relation $\Pi_R^{\m\n}(q)=\(g^{\m\n}-q^\m q^\n/q^2\){\tilde \Pi_R}(q)$. Within PT one should propose that ${\tilde \Pi_R}(q)\ll q^2$, while the potential $A_0$ should be considered within the expansion  $A_0(q)=A^{(0)}(q)+A^{(1)}(q)+\ldots$. The classical part of the potential is determined from the equation
\beq \label{q4}
q^2 A^{(0)}(q)=\r(q) \ ,
\eeq
while its first quantum correction --- from equation
\beq \label{q5}
q^2 A^{(1)}(q)={\tilde \Pi_R}(q)A^{(0)}(q)  \ .
\eeq
The r.h.s. in (\ref{q5}) is just the induced vacuum charge density $\r^{(1)}_{VP}(q)$ to the first order of PT. Then in the coordinate representation (up to multipliers like $2\pi$)
\beq \label{q6}
\r^{(1)}_{VP}(x)=\int \! dy \ {\tilde \Pi_R}(x-y) A^{(0)}(y)=\int \! dq \ {\tilde \Pi_R}(q) A^{(0)}(q) \mathrm{e}^{iqx} \ .
\eeq
Upon integrating this equality over all  $x$, one obtains the total induced charge $Q^{(1)}_{VP}$ to the first order of PT
\beq \label{q7}
Q^{(1)}_{VP}=\int \! dq \ \d(q) {\tilde \Pi_R}(q) A^{(0)}(q) \ .
\eeq
 And since in PT there holds the renormalization condition ${\tilde \Pi_R}(q) \sim q^2$ for $q^2 \to 0$, there follows from  (\ref{q7}) that  $Q^{(1)}_{VP}=0$ subject to condition, that the singularity of the Fouriet-transform of the external potential $A^{(0)}(q)$ for $q\to 0$ is weaker, than $1/q^2$. For the Coulomb potentials like (\ref{1.1}) in 1+1 D  the singularity of $A^{(0)}(q)$ is just a logarithmic one, hence  $Q^{(1)}_{VP}\equiv 0$. In 2+1 and 3+1 D the singularity of $A^{(0)}(q)$ is already stronger, $1/| \vec q|$ and $1/{\vec q}^2$ correspondingly, but  thereat appears an additional factor $|\vec q|^{D-1}$ from the integration measure, which compensates these singularities already by itself. Therefore for the  Coulomb potentials like (\ref{1.00}) in all the three spatial dimensions there follows $Q^{(1)}_{VP}\equiv 0$ to the first order of PT. It is indeed this simple reasoning that was implied in Section 2 under the statement ``in this case the relation  (\ref{2.7}) is the direct consequence of the renormalization condition''.

However, beyond the first-order PT and especially in the whole subcritical region $Z<Z_{cr,1}$, when in presence of negative discrete levels the dependence of $\r_{VP}$ on the external field cannot be described by the power series (\ref{3.211}) any more, the check of the zero value of $Q^{ren}_{VP}$ requires a sufficiently more detailed consideration, that in 1+1 D, where it could be performed almost completely in a purely analytical form \cite{davydov2017}, but in 2+1 and 3+1 requires for an additional numerical doings.

In 2+1 D this check could be most efficiently performed in the next way. In the first step, let us take into account the uniform convergence of the partial series in $m_j$ with respect to  $r$, considered as an external parameter in (\ref{q2}) (see Appendix C).  Then it is possible to insert the integral under the sign of the sum in the expression for $Q^{(3+)}_{VP}$ \beq \label{q8}
\bal
&Q^{(3+)}_{VP}= \sum_{m_{j}=1/2,\,3/2,..} Q^{(3+)}_{VP,|m_{j}|} \ , \qquad Q^{(3+)}_{VP,|m_{j}|}= 4 \pi \int\limits_{0}^{\infty} \! r\, dr\, \r^{(3+)}_{VP, |m_{j}|}(r) \ .
\eal
\eeq
Proceeding further, from the explicit form of the vacuum density $\r^{(3+)}_{VP, |m_{j}|}(r)$, given in (\ref{3.23}),  for $Q^{(3+)}_{VP,|m_{j}|}$  one obtains the following expression
\beq \label{q9}
\bal
Q^{(3+)}_{VP,|m_{j}|}&= 2 |e| \[\sum_{m_j=\pm|m_j|} \sum_{-1\leq \e_{n,m_j} < 0}1\right.\\
&\left.+\frac{1}{\pi}\int\limits_{0}^{\infty} \! r\, dr \ \int\limits_{0}^{\infty}dy\, \mathrm{Re}\(\tr G_{|m_j|}(r,r;iy)-2\, \tr G^{(1)}_{m_j}(r;iy)\)\] \ ,
\eal
\eeq

Now let us show that in the subcritical region $Q^{(3+)}_{VP,|m_j|}$ for all  $m_j$ vanish exactly, and so the total induced charge $Q^{ren}_{VP}$  does the same. The most direct method here is the straightforward numerical calculation of the double integral in (\ref{q9}). However, it turns out to be a sufficiently time-consuming numerical task. As an alternative way, it is possible to take advantage from the following circumstances. First, as it was stated above in the Section 4, it suffices to verify the disappearance of $Q_{VP}^{ren}$ not for the whole subcritical region, but only in absence of negative discrete levels. The second point is the observation, that if the integrals over $dr$ in (\ref{q9}) become the internal ones, they could be calculated analytically by means of Ref.~\cite{prud2002}. For these purposes let us exchange the sequence of integrations in (\ref{q9}), inserting the intermediate regularization  of $ \int \! dr$ at the upper limit, otherwise the exchange of integrations and especially inserting $ \int \! dr$ under the sign of derivative with respect to $Q$ is not allowed, since $ \int \! r\, dr \ \tr G_{|m_j|}(r,r;iy)$ is logarithmically divergent for large $r$ due to the asymptotics (\ref{3.22}) of $\tr G_{|m_j|}(r,r;iy)$.

  With account of these circumstances it suffices to deal with $Q^{(3+)}_{VP,|m_j|}$ in the following  form
\beq \label{q10}
\bal
Q^{(3+)}_{VP,|m_{j}|}&=  \frac{2 |e|}{\pi} \int\limits_{0}^{\infty} \! dy \ \lim_{R_1\to \infty}\mathrm{Re}\[\int\limits_{0}^{R_1} \! r\, dr \ \tr G_{|m_j|}(r,r;iy)\right. \\
&\left.-2\, Q \(\frac{\pd}{\pd Q} \int\limits_{0}^{R_1} \! r\, dr\, \tr G_{m_j}(r,r;iy)\)_{Q=0} \] \ .
\eal
\eeq
Let us consider now   $ \int_0^{R_1} \! r\, dr\, \tr G_{m_j}(r,r;iy)$ in (\ref{q10}). Taking into account the following indefinite integrals \cite{prud2002}
\beq \label{q10.1}
\bal
&\int \frac{1}{x}U_{\r,\s}(x)\hat{U}_{\r,\s}(x) dx=-U_{\r,\s}(x) \frac{\pd}{\pd \r}\hat{U}_{\r,\s}'(x)+U_{\r,\s}'(x)\frac{\pd}{\pd \r} \hat{U}_{\r,\s}'(x) \ , \\
&\int \frac{1}{x} W_{\m,\s}(x) W_{\r,\s}(x) dx=\frac{1}{\m-\r}\(W_{\r,\s}'(x)W_{\m,\s}(x)-W_{\r,\s}(x)W_{\m,\s}'(x)\) \ , \\
&\int \(\frac{b^2-a^2}{4}+\frac{\m a - \r b}{x}+\frac{\n^2-\s^2}{x^2}\)U_{\m,\n}(ax)\hat{U}_{\r,\s}(bx) dx\\
&=U_{\m,\n}(ax) \hat{U}'_{\r,\s}(bx) -U_{\m,\n}'(ax) \hat{U}_{\r,\s}(bx) \ ,
\eal
\eeq
where $U_{\r,\s}(x), \hat{U}_{\r,\s}(x) = M_{\r,\s}(x) \  \text{or} \ W_{\r,\s}(x)$, the prime stands for the derivative with respect to the argument, and the explicit form of $\tr G_{m_j}(r,r;iy)$, which is given in (\ref{3.11}), one obtains
\beq \label{q11}
\bal
&\int\limits_{0}^{R_1} \! r\, dr\, \tr G_{m_j}(r,r;iy)\\
&=\left.\frac{1}{\[\cI,\cK\]}\(J_{1}-\frac{\[\cK,\cW\]_R}{\[\cI,\cW\]_R}J_{2}\)\right|_{\e=i y}+\left.\frac{1}{\[\cM,\cW\]}\(J_{3}-\frac{\[\cI,\cM\]_R}{\[\cI,\cW\]_R}J_4\)\right|_{\e=iy} \ ,
\eal
\eeq
where
\beq \label{q12}
\bal
J_{1}&=\int\limits_{0}^{R} \! r\, dr\, \( \cI_{1} \cK_{1} + \cI_{2} \cK_{2} \)\\
&=(1-\e-V_{0})^2 \left\lbrace\frac{R^2}{2}  \[\(\frac{\(m_{j}+1/2\)^2}{\(\x R\)^2 }+1\) I_{| m_{j}+1/2| }(\x R) K_{| m_{j}+1/2| }(\x R )\right.\right. \\
&\left.\left.+\frac{1}{4}\(I_{| m_{j}+1/2| -1}( \x R)+I_{| m_{j}+1/2| +1}(\x R )\) \(K_{| m_{j}+1/2| -1}(\x R)+K_{| m_{j}+1/2| +1}(\x R)\)\]\right.\\
&\left.-\frac{| m_{j}+1/2| }{2 \x^2}\right\rbrace-\x^2 \left\lbrace\frac{R^2}{2}  \[\(\frac{\(m_{j}-1/2\)^2}{\(\x R \)^2}+1\) I_{| m_{j}-1/2| }(\x R) K_{| m_{j}-1/2| }(\x R)\right.\right.\\
&\left.\left.+\frac{1}{4}\(I_{| m_{j}-1/2| -1}(\x R)+I_{| m_{j}-1/2| +1}(\x R)\) \(K_{| m_{j}-1/2| -1}(\x R) + K_{| m_{j}-1/2| +1}(\x R)\)\]\right.\\
&\left.-\frac{| m_{j}-1/2| }{2 \x^2}\right\rbrace \ ,
\eal
\eeq
\beq \label{q13}
\bal
J_{2}&=\int\limits_{0}^{R} \! r\, dr\, \( \cI^{2}_{1} + \cI^{2}_{2} \)\\
&= \frac{r^2}{2} \left\lbrace\(V_{0}+(\e -1)\)^2 \[I_{| m_{j}+1/2| }\(\x R\)^2-I_{| m_{j}+1/2| -1}\(\x R\) I_{| m_{j}+1/2| +1}\(\x R\)\]\right.\\
&\left.+\x^2 \[I_{| m_{j}-1/2| }\(\x R\)^2-I_{| m_{j}-1/2| -1}\(\x R\) I_{| m_{j}-1/2| +1}\(\x R\)\]\right\rbrace \ ,
\eal
\eeq
\beq \label{q14}
\bal
&J_{3}=\int\limits_R^{R_{1}} \! r\, dr\, \( \cM_{1} \cW_{1} + \cM_{2} \cW_{2}\)=\overline{J_{3}}(R_{1})-\overline{J_{3}}(R) \ , \\
&\overline{J_3}(r)=-2 (\e +1) \left\lbrace\(m_{j}+\frac{Q}{\g}\)\right. \\
&\left.\times \[\(\(\frac{1}{2}-\frac{\n+1/2}{2 \g r}\) M_{\n+1/2,s}(2 \g r)+\frac{s+\n+1}{2 \g r} M_{\n+3/2,s}(2 \g r)\) \left.\frac{\pd}{\pd \m_{1}}W_{\m_{1},s}(2 \g r)\right|_{\m_{1}=\n+1/2}\right.\right.\\
&\left.\left.-M_{\n+1/2,s}(2 \g r) \left.\frac{\pd}{\pd \m_{1}}\(\(\frac{1}{2}-\frac{\m_{1}}{2 \g r}\) W_{\m_{1},s}(2 \g r)-\frac{W_{\m_{1}+1,s}(2 \g r)}{2 \g r}\)\right|_{\m_{1}=\n+1/2}\]\right.\\
&\left.+(s-\n) \left\lbrace\(\frac{Q}{\g}-m_{j}\)\right.\right.\\
&\left.\left.\times\[\(\(\frac{1}{2}-\frac{\n-1/2}{2 \g r}\) M_{\n-1/2,s}(2 \g r)+\frac{s+\n}{2 \g r} M_{\n+1/2,s}(2 \g r)\)\left.\frac{\pd}{\pd \m_{2}}W_{\m_{2},s}(2 \g r)\right|_{\m_{2}=\n-1/2} \right.\right.\right.\\
&\left.\left.\left.-M_{\n-1/2,s}(2 \g r) \left.\frac{\pd }{\pd \m_{2}}\(\(\frac{1}{2}-\frac{\m_{2}}{2 \g r}\) W_{\m_{2},s}(2 \g r)-\frac{W_{\m_{2}+1,s}(2 \g r)}{2 \g r}\)\right|_{\m_{2}=\n-1/2}\]\right.\right. \\
&\left.\left.-\e  \[M_{\n-1/2,s}(2 \g r) \(\(\frac{1}{2}-\frac{\n+1/2}{2 \g r}\) W_{\n+1/2,s}(2 \g r)-\frac{W_{\n+3/2,s}(2 \g r)}{2 \g r}\)\right.\right.\right.\\
&\left.\left.\left.-\(\(\frac{1}{2}-\frac{\n-1/2}{2 \g r}\) M_{\n-1/2,s}(2 \g r)+\frac{s+\n}{2 \g r}M_{\n+1/2,s}(2 \g r)\) W_{\n+1/2,s}(2 \g r)\]\right\rbrace\right.\\
&\left.+\e  \(\frac{Q^2}{\g^2}-m_{j}^2\) \[M_{\n+1/2,s}(2 \g r) \(\(\frac{1}{2}-\frac{\n-1/2}{2 \g r}\) W_{\n-1/2,s}(2 \g r)-\frac{W_{\n+1/2,s}(2 \g r)}{2 \g r}\)\right.\right.\\
&\left.\left.-\(\(\frac{1}{2}-\frac{\n+1/2}{2 \g r}\) M_{\n+1/2,s}(2 \g r)+\frac{s+\n+1 }{2 \g r}M_{\n+3/2,s}(2 \g r)\) W_{\n-1/2,s}(2 \g r)\]\right\rbrace \ ,
\eal
\eeq
\beq \label{q15}
\bal
&J_{4}=\int\limits_R^{R_{1}} \! r\, dr\, \(\cW^{2}_{1} + \cW^{2}_{2}\)=\overline{J_{41}}(R_{1})-\overline{J_{41}}(R)+\overline{J_{42}}(R_{1})-\overline{J_{42}}(R) \ , \\
&\begin{bmatrix}
\overline{J_{41}}(r) \\
\overline{J_{42}}(r)
\end{bmatrix}=\begin{bmatrix}
(1+\epsilon)^2 \\
\g^2
\end{bmatrix} \left\lbrace\(m_{j}-\frac{Q}{\g}\)^2 \right.\\
&\left.\times\[\(\(\frac{1}{2}-\frac{\m_{1}}{2 \g  r }\) W_{\m_{1},s}(2 \g r)-\frac{W_{\m_{1}+1,s}(2 \g r)}{2 \g  r }\) \frac{\pd}{\pd \m_{1}} W_{\m_{1},s}(2 \g r)\right.\right.\\
&\left.\left.-W_{\m_{1},s}(2 \g r) \left.\frac{\pd }{\pd \m_{1}}\(\(\frac{1}{2}-\frac{\m_{1}}{2 \g  r }\) W_{\m_{1},s}(2 \g r)-\frac{W_{\m_{1}+1,s}(2 \g r)}{2 \g  r }\)\]\right|_{\m_{1}= \n-1/2}\right. \\
&\left.+\[\(\(\frac{1}{2}-\frac{\m_{2}}{2 \g  r }\) W_{\m_{2},s}(2 \g r)-\frac{W_{\m_{2}+1,s}(2 \g r)}{2 \g  r }\) \frac{\pd}{\pd \m_{2}} W_{\m_{2},s}(2 \g r)\right.\right.\\
&\left.\left.-W_{\m_{2},s}(2 \g r) \left.\frac{\pd }{\pd \m_{2}}\(\(\frac{1}{2}-\frac{\m_{2}}{2 \g  r }\) W_{\m_{2},s}(2 \g r)-\frac{W_{\m_{2}+1,s}(2 \g r)}{2 \g  r }\)\]\right|_{\m_{2}=\n+1/2}\right. \\
&\left.\begin{bmatrix}
+\\
-
\end{bmatrix} 2 \(m_{j}-\frac{Q}{\g}\) \right.\\
&\left.\times\[\(\(\frac{1}{2}-\frac{\n+1/2}{2 \g  r }\) W_{\n+1/2,s}(2 \g r)-\frac{W_{\n+3/2,s}(2 \g r)}{2 \g  r }\) W_{\n-1/2,s}(2 \g r)\right.\right.\\
&\left.\left.-\(\(\frac{1}{2}-\frac{\n-1/2}{2 \g  r }\) W_{\n-1/2,s}(2 \g r)-\frac{W_{\n+1/2,s}(2 \g r)}{2 \g  r }\) W_{\n+1/2,s}(2 \g r)\]\right\rbrace \ .
\eal
\eeq
In (\ref{q12})-(\ref{q15}) the notations are the same as introduced earlier in Section 3. In the next step, the limit $R_1 \to \inf$ is calculated. In this limit the linear in $Q$ logarithmically divergent terms, originating from the asymptotics (\ref{3.22}), cancel each other, whence
\beq \label{q16}
\bal
&\lim_{R_{1} \to \infty} \mathrm{Re}\[\int\limits_{0}^{R_{1}}\! r\, dr\, \tr G_{|m_{j}|}(r,r;i y) - 2 Q \(\frac{\pd}{\pd Q}\int\limits_{0}^{R_{1}} \! r\, dr\, \tr G_{m_{j}}(r,r;iy)\)_{Q=0}\] \\
&=\mathrm{Re}\[ \frac{1}{\[\cI,\cK\]}\(J_{1}-\frac{\[\cK,\cW\]_{R}}{\[\cI,\cW\]_{R}} J_{2}\) + \frac{1}{\[\cI,\cK\]}\(J_{1}-\frac{\[\cK,\cW\]_{R}}{\[\cI,\cW\]_{R}} J_{2}\)_{ m_j\to - m_ j} \right. \\
&\left. -\frac{1}{\[\cM,\cW\]}\(\overline{J_{3}}(R)-\frac{\[\cI,\cM\]_{R}}{\[\cI,\cW\]_{R}} \(\overline{J_{41}}(R)+\overline{J_{42}}(R)\) \)\right.\\
&\left.-\frac{1}{\[\cM,\cW\]}\(\overline{J_{3}}(R)-\frac{\[\cI,\cM\]_{R}}{\[\cI,\cW\]_{R}} \(\overline{J_{41}}(R)+\overline{J_{42}}(R)\) \)_{m_{j} \to -m_{j}} \right. \\
&\left. -2 Q \[\frac{\pd }{\pd Q}\(\frac{1}{\[\cI,\cK\]}\(J_{1}-\frac{\[\cK,\cW\]_{R}}{\[\cI,\cW\]_{R}} J_{2}\)\right.\right.\right.\\
&\left.\left.\left. -\frac{1}{\[\cM,\cW\]}\(\overline{J_{3}}(R)-\frac{\[\cI,\cM\]_{R}}{\[\cI,\cW\]_{R}} \(\overline{J_{41}}(R)+\overline{J_{42}}(R)\) \)  \) \]_{Q=0} \]_{\e=iy} \ .
\eal
\eeq
In (\ref{q16}) the derivatives with respect to $Q$ are not shown explicitly due to their cumbersome form.

As a result, there remains only a single numerical integration over $dy$, which despite the complexity of the integrand does not already pose any problems, since the integral is definitely convergent. Namely, the asymptotical behavior of the integrand is estimated as $\sim 1/|y|^{5}$ (see Appendix C below). Therefore such integration can be performed via standard numerical recipes, and in this way by means  of (\ref{q10})-({\ref{q16}) one can verify that in the subcritical region in absence of negative   levels the total induced charge vanishes. More concretely, we have checked by explicit calculations with WorkingPrecision$\to 100$ and PrecisionGoal$\to 15$ that for $Z=50$, when there are no negative discrete levels yet, $ Q^{(3+)}_{VP,1/2}=2.5582 \times 10^{-30} |e| \ , \  Q^{(3+)}_{VP,3/2}=2.4007 \times 10^{-41} |e|$, while the other partial charges $Q^{(3+)}_{VP,|m_j|}$ with higher $|m_j|$ decrease further according to  the law $|m_j|^{-3}$. These results look quite convincing for the assertion that in the subcritical    region $Q^{ren}_{VP}\equiv 0$. In the same way it is possible to verify that in the overcritical   $Z >Z_{cr,1}$ the total vacuum charge  $Q^{ren}_{VP}$  is equal to an integer number of $(-2|e|)$ in dependence on the number of levels, which have dived into the lower continuum, and with account of their degeneracy.

\section{Verifying the  Uniform Convergence of the Integral $\int \! dy \ \mathrm{Re}\[ \tr G_{|m_j|}(r,r;iy)-2\, \tr G^{(1)}_{m_j}(r;i y)\]$}

Here it will be shown  that the integral $\int \! dy \ \mathrm{Re}\[ \tr G_{|m_j|}(r,r;iy)-2\, \tr G^{(1)}_{m_j}(r;i y)\]$, that defines the main component of $\r_{VP,|m_j|}^{(3+)}(r)$ in (\ref{3.23}), converges uniformly with respect to $m_j$ and $r$. For these purposes $\tr G_{m_{j}}$ should be represented as follows:
\beq \label{c1}
\tr G_{m_{j}}(r,r;\e) = \tt(R-r) \tr G_{m_{j}}^{in}(r,r;\e) + \tt(r-R) \tr G_{m_{j}}^{out}(r,r;\e) \ , \eeq
where
\beq \label{c2}
\bal
&\tr G_{m_{j}}^{in}(r,r;\e)=\tr G_{m_{j}}^{0,in}(r,r;\e)+ \tr \D G_{m_j}^{in}(r,r;\e) \ , \\
&\tr G_{m_{j}}^{0,in}(r,r;\e)=\frac{1}{\[\cI,\cK\]}\(\cI_{1} \cK_{1} + \cI_{2} \cK_{2} \) \ ,\\
&\tr \D G_{m_j}^{in}(r,r;\e)=-\frac{1}{\[\cI,\cK\]}\frac{\[\cK,\cW\]_{R}}{\[\cI,\cW\]_{R}}\(\cI_{1}^{2}+\cI_{2}^{2}\) \ ,
\eal
\eeq
and
\beq \label{c3}
\bal
&\tr G_{m_{j}}^{out}(r,r;\e)=\tr G_{m_{j}}^{0,out}(r,r;\e)+ \tr \D G_{m_j}^{out}(r,r;\e) \ , \\
&\tr G_{m_{j}}^{0,out}(r,r;\e)=\frac{1}{\[\cM,\cW\]}\(\cM_{1} \cW_{1} + \cM_{2} \cW_{2} \) \ , \\
&\tr \D G_{m_j}^{in}(r,r;\e)=-\frac{1}{\[\cM,\cW\]}\frac{\[\cI,\cM\]_{R}}{\[\cI,\cW\]_{R}}\(\cW_{1}^{2}+\cW_{2}^{2}\) \ .
\eal
\eeq
In (\ref{c2})-(\ref{c3}) the notations, introduced earlier in the main text (\ref{3.11}), are used.

Now let us consider more thoroughly, up to $\mathrm{O}\(1/\e^{5}\)$ inclusively, the asymptotics of $\tr G_{m_{j}}$ for $\e$ on  the arcs of the large circle in the upper half-plane $C_{1}$ and $C_{2}$ (Fig.~1) ($|\e|\to \infty, 0< \mathrm{Arg}\, \e<\pi$). The corresponding asymptotics on the arcs of the large circle in the lower half-plane could be obtained then from general properties of $\tr G_{m_{j}}$ (33).

The asymptotics of $\tr G_{m_{j}}$ for $0< r < R$ has the following form (about the vicinity of the point $r=0$ see below)
\beq \label{c4}
\bal
&\tr G^{0,in}_{m_{j}}(r,r;\e) \to C^{in}_{0}(r) + \frac{C^{in}_{2}(r)}{\e^{2}} + \frac{C^{in}_{3}(r)}{\e^{3}} + \frac{C^{in}_{4}(r)}{\e^{4}} + \frac{C^{in}_{5}(r)}{\e^{5}} + \mathrm{O\(\frac{1}{|\e|^{6}}\)} \ , \\
&C^{in}_{0}(r)=\frac{i}{r}\ , \qquad C^{in}_{2}(r)=\frac{i}{2 r}\(\frac{m_{j}^{2}}{r^2}+1\) \ , \\
&C^{in}_{3}(r)=-\frac{i}{r^2} \(\frac{m_{j}^2}{r} V_{0} +\frac{m_{j}}{2 r}+r V_{0}\) \ , \\
&C^{in}_{4}(r)= \frac{3 i}{2 r^3}\(\frac{\(m_{j}^2-1\right) m_{j}^2}{4 r^2}+\frac{m_{j}^2}{2}+m_{j}^{2} V_{0}^{2} +m_{j} V_{0}+r^2 V_{0}^2+\frac{r^2}{4}\right)\ , \\
&C^{in}_{5}(r)= \frac{3 i}{r^4} \(\frac{-2 m_{j}^4 V_{0}-m_{j}^3+2 m_{j}^2 V_{0}+m_{j}}{4 r}\right.\\
&\left.-r \(\frac{2}{3}m_{j}^2 V_{0}^3+m_{j}^2 V_{0}+m_{j} V_{0}^2+\frac{m_{j}}{4}\)-r^3 \(\frac{2}{3}V_{0}^3+\frac{V_{0}}{2}\)\) \ .
\eal
\eeq
\beq \label{c5}
\bal
&\tr \D G^{in}_{m_{j}}(r,r;\e) \to \mathrm{e}^{-2 \x \(R-r\)}\[\frac{D^{in}_{4}(r)}{\e^{4}}+\frac{D^{in}_{5}(r)}{\e^{5}}+\mathrm{O}\(\frac{1}{|\e|^6}\)\] \ , \\
&D^{in}_{4}(r)=-\frac{1}{4 r R^2}\(\frac{m_{j}^2 Q}{r R}-\frac{i m_{j} Q}{r}+\frac{i m_{j} Q}{R}+Q\) \ , \\
&D^{in}_{5}(r)=-\frac{1}{r R^2}\(-\frac{i m_{j}^4 Q}{4 r^2 R}+\frac{i m_{j}^4 Q}{4 r R^2}-\frac{m_{j}^3 Q}{4 r^2}+\frac{m_{j}^3 Q}{2 r R}-\frac{m_{j}^3 Q}{4 R^2}+\frac{m_{j}^2 Q^2}{r R^2}+\frac{i m_{j}^2 Q}{8 r^2 R}\right.\\
&\left.-\frac{5 i m_{j}^2 Q}{8 r R^2}-\frac{i m_{j}^2 Q}{4 r}+\frac{i m_{j}^2 Q}{4 R}- \frac{i m_{j} Q^2}{r R}+\frac{i m_{j} Q^2}{R^2}+\frac{m_{j} Q}{8 r^2}-\frac{m_{j} Q}{4 r R}\right.\\
&\left.+\frac{5 m_{j} Q}{8 R^2}+\frac{Q^2}{R}-\frac{i Q}{4 R}\) .
\eal
\eeq
It follows from (\ref{c4}) and  (\ref{c5}) with account of $\mathrm{Re}\, \x>0$ (22) that the asymptotics of $\tr G_{m_{j}}$ for $r < R$ is defined via asymptotics of $\tr G^{0,in}_{m_{j}}$. At the same time, for $r \to R$ one should  take into account that the contribution from $\tr \D G^{in}_{m_{j}}$ becomes non-zero.

The asymptotics of $\tr G_{m_{j}}$ for $r > R$ reveals the same structure:
\beq \label{c6}
\bal
&\tr G^{0,out}_{m_{j}}(r,r;\e) \to C^{out}_{0}(r) + \frac{C^{out}_{2}(r)}{\e^{2}} + \frac{C^{out}_{3}(r)}{\e^{3}} + \frac{C^{out}_{4}(r)}{\e^{4}} + \frac{C^{out}_{5}(r)}{\e^{5}} + \mathrm{O\(\frac{1}{|\e|^{6}}\)} \ , \\
&C^{out}_{0}(r)=\frac{i}{r}\ , \qquad C^{out}_{2}(r)=\frac{i}{2 r}\(\frac{m_{j}^{2}}{r^2}+1\) \ , \\
&C^{out}_{3}(r)=-\frac{i}{r^2} \(\frac{m_{j}^2}{r} \frac{Q}{r} +\frac{m_{j}}{2 r}+Q\) \ , \\
&C^{out}_{4}(r)= \frac{3 i}{2 r^3}\(\frac{\(m_{j}^2-1\right) m_{j}^2}{4 r^2}+\frac{m_{j}^2}{2}+m_{j}^{2} \(\frac{Q}{r}\)^{2}+ m_{j} \frac{Q}{r}+Q^2+\frac{r^2}{4}\right)\ , \\
&C^{out}_{5}(r)= \frac{3 i}{r^4} \[\frac{-2 m_{j}^4 Q/r-m_{j}^3+10/3 m_{j}^2 Q/r+m_{j}}{4 r}\ -\right.\\
&\left. - r \(\frac{2}{3}m_{j}^2 \(\frac{Q}{r}\)^3+m_{j}^2 \frac{Q}{r}+m_{j} \(\frac{Q}{r}\)^2+\frac{m_{j}}{4}-\frac{Q}{6 r}\)-r^3 \(\frac{2}{3}\(\frac{Q}{r}\)^3+\frac{Q}{2 r}\)\] \ .
\eal
\eeq
\beq \label{c7}
\bal
&\tr \D G^{out}_{m_{j}}(r,r;\e) \to \mathrm{e}^{-2 \g \(r-R\)}\(\frac{R}{r}\)^{i 2 Q}\[\frac{D^{out}_{4}(r)}{\e^{4}}+\frac{D^{out}_{5}(r)}{\e^{5}}+\mathrm{O}\(\frac{1}{|\e|^6}\)\] \ , \\
&D^{out}_{4}(r)=-\frac{1}{4 r R^2}\(\frac{m_{j}^2 Q}{r R}+\frac{i m_{j} Q}{r}-\frac{i m_{j} Q}{R}+Q\) \ , \\
&D^{out}_{5}(r)=-\frac{1}{r R^2}\(\frac{i m_{j}^4 Q}{4 r^2 R}-\frac{i m_{j}^4 Q}{4 r R^2}-\frac{m_{j}^3 Q}{4 r^2}+\frac{m_{j}^3 Q}{2 r R}-\frac{m_{j}^3 Q}{4 R^2}+\frac{m_{j}^2 Q^2}{4 r^2 R}\right.\\
&\left.+\frac{3 m_{j}^2 Q^2}{4 r R^2}-\frac{i m_{j}^2 Q}{8 r^2 R}+\frac{5 i m_{j}^2 Q}{8 r R^2}+\frac{i m_{j}^2 Q}{4 r}-\frac{i m_{j}^2 Q}{4 R}+\frac{i m_{j} Q^2}{4 r^2}+\frac{i m_{j} Q^2}{2 r R}\right.\\
&\left.-\frac{3 i m_{j} Q^2}{4 R^2}+\frac{m_{j} Q}{8 r^2}-\frac{m_{j} Q}{4 r R}+\frac{5 m_{j} Q}{8 R^2}+\frac{Q^2}{4 r}+\frac{3 Q^2}{4 R}+\frac{i Q}{4 R}\) .
\eal
\eeq
From (\ref{c6}) and  (\ref{c7}) with account of $\mathrm{Re}\, \g>0$ (24) one finds that the asymptotics of $\tr G_{m_{j}}$ for $r > R$ is defined by the asymptotics of $\tr G^{0,out}_{m_{j}}$, whereas for $r \to R$, on the contrary, the contribution of $ \tr \D G^{out}_{m_{j}}$ cannot be neglected.

 Now let us verify that with account of the contributions of $\tr \D G^{in,out}_{m_{j}}$ the asymptotics of $\tr G^{in}_{m_{j}}(R,R;\e)$ and $\tr G^{out}_{m_{j}}(R,R;\e)$ actually coincide. Indeed, from (\ref{c4})-(\ref{c7}) at $r=R$ one obtains
\beq \label{c8}
\bal
&\tr G^{in}_{m_{j}}(R,R;\e) \to C^{in}_{0}(R) + \frac{C^{in}_{2}(R)}{\e^{2}} + \frac{C^{in}_{3}(R)}{\e^{3}} + \frac{C^{in}_{4}(R)+D^{in}_{4}(R)}{\e^{4}} \\
&+ \frac{C^{in}_{5}(R)+D^{in}_{5}(R)}{\e^{5}} + \mathrm{O\(\frac{1}{|\e|^{6}}\)} \ , \\
&\tr G^{out}_{m_{j}}(R,R;\e) \to C^{out}_{0}(R) + \frac{C^{out}_{2}(R)}{\e^{2}} + \frac{C^{out}_{3}(R)}{\e^{3}} + \frac{C^{out}_{4}(R)+D^{out}_{4}(R)}{\e^{4}} \\
&+ \frac{C^{out}_{5}(R)+D^{out}_{5}(R)}{\e^{5}} + \mathrm{O\(\frac{1}{|\e|^{6}}\)} \ ,
\eal
\eeq
and the following relations between the coefficients of the in- and out-expansions
\beq \label{c9}
\bal
&C^{in}_{0}(R)=C^{out}_{0}(R) \ , \qquad C^{in}_{2}(R)=C^{out}_{2}(R) \ , \qquad C^{in}_{3}(R)=C^{out}_{3}(R) \ , \\
&C^{in}_{4}(R)=C^{out}_{4}(R) \ , \qquad C^{in}_{5}(R)-C^{out}_{5}(R)=-Q\(\frac{m_j^2}{3 R^2}+\frac{1}{6}\) \ ,\\
&D^{in}_{4}(R)= D^{out}_{4}(R) \ , \qquad D^{in}_{5}(R) - D^{out}_{5}(R)=Q\(\frac{m_j^2}{3 R^2}+\frac{1}{6}\) \ .
\eal
\eeq
From (\ref{c8}) and (\ref{c9}) there follows that at $r=R$ the asymptotics of $\tr G_{m_{j}}$ is continuous and takes the form (\ref{c8}).

Proceeding further and making use of the general properties of $\tr G_{m_{j}}$ (33), one finds the asymptotics of $\tr G_{m_{j}}$ for $\e$ on the arcs of the large circle  in the lower half-plane $C_{3}$ and $C_{4}$ (Fig.~1) ($|\e|\to \infty, -\pi< \mathrm{Arg}\, \e<0$). As a result, the asymptotics of $\mathrm{Re} \tr G_{|m_j|}(r,r;iy)$ for $|y| \to \infty$, considered in Section 3 in terms of $ \tr G_{|m_j|}(r,r;\e)$ up to $\mathrm{O}\(1/\e^{3}\)$ only,   takes now the following form:\\
for $r<R$
\beq \label{c10}
\bal
\frac{2}{r^2|y|^3}\(\frac{m_{j}^{2}}{r}V_{0}+r V_{0}\)&+\frac{6}{r^4 |y|^5}\[-\frac{m_{j}^{4}}{2r}V_{0}+\frac{m_{j}^{2}}{2r}V_{0}-r\(\frac{2}{3}m_{j}^{2}V_{0}^{3}+m_{j}^{2}V_{0}\)\right.\\
&\left.-r^{3}\(\frac{2}{3}V_{0}^{3}+\frac{V_{0}}{2}\)\]+\mathrm{O}\(\frac{1}{|y|^{7}}\) \ ;
\eal
\eeq
at $r=R$
\beq \label{c11}
\bal
&\frac{2}{R^2|y|^3}\(\frac{m_{j}^{2}}{R}V_{0}+Q\)-\frac{1}{2 R^3 y^4}\(\frac{m_{j}^{2}}{R}V_{0}+Q\)+ \frac{6}{R^4 |y|^5}\[-\frac{m_{j}^4}{2 R}V_{0}+\frac{2 m_{j}^2}{3R}V_{0}\right.\\
&\left.-R\(\frac{2}{3} m_{j}^2V_{0}^{3}+m_{j}^2 V_{0}-\frac{V_{0}}{12}\)-R^{3}\(\frac{2}{3} V_{0}^3+\frac{V_{0}}{2}\)\]+\mathrm{O}\(\frac{1}{|y|^{6}}\) \ ;
\eal
\eeq
for $r>R$
\beq \label{c12}
\bal
&\frac{2}{r^2|y|^3}\(\frac{m_{j}^{2}}{r}\frac{Q}{r}+Q\)+\frac{6}{r^4 |y|^5}\[-\frac{m_{j}^{4}}{2r}\frac{Q}{r}+\frac{5 m_{j}^{2}}{6 r}\frac{Q}{r}-r\(\frac{2}{3}m_{j}^2 \(\frac{Q}{r}\)^3+m_{j}^2 \frac{Q}{r}-\frac{Q}{6 r}\)\right.\\
&\left.-r^3 \(\frac{2}{3}\(\frac{Q}{r}\)^3+\frac{Q}{2 r}\)\] + \mathrm{O}\(\frac{1}{|y|^{7}}\) \ .
\eal\eeq
Let us specially note that from (\ref{c10})-(\ref{c12}) there might appear an impression, that they do not provide the continuity at $r=R$. However, as it has been already mentioned above, for $r \to R$ one should take into account in the asymptotics of $\tr G^{in}_{m_{j}}(r,r;iy)$ and $\tr G^{out}_{m_{j}}(r,r;iy)$ the non-vanishing contributions from $\tr \D G^{in}_{m_{j}}$ and $\tr \D G^{out}_{m_{j}}$. Namely, on account of them the asymptotics $\mathrm{Re}\tr G^{in}_{|m_{j}|}(r,r;iy)$ for $|y| \to \inf$ can be represented as follows
\beq \label{com1}
\bal
&\mathrm{Re}\tr G^{in}_{|m_{j}|}(r,r;iy) \to \frac{2}{r^2|y|^3}\(\frac{m_{j}^{2}}{r}V_{0}+r V_{0}\)+\frac{6}{r^4 |y|^5}\[-\frac{m_{j}^{4}}{2r}V_{0}+{\frac{m_{j}^{2}}{2 r}V_{0}}\right. \\
&\left.-r\(\frac{2}{3}m_{j}^{2}V_{0}^{3}+m_{j}^{2}V_{0}\)-r^{3}\(\frac{2}{3}V_{0}^{3}+\frac{V_{0}}{2}\)\]+\mathrm{O}\(\frac{1}{|y|^{7}}\) + \mathrm{Re}\(\left.\mathrm{e}^{-2 \x \(R-r\)}\right|_{\e=i y}\right. \\
&\left. \times \[\frac{D_{4}^{in}(r)+D_{4}^{in}(r)|_{m_j\to - m_j}}{y^{4}}+\frac{D_{5}^{in}(r)+D_{5}^{in}(r)|_{m_j \to - m_j }}{i y^{5}}+\mathrm{O}\(\frac{1}{|y|^{6}}\) \]\) \ .
\eal
\eeq
Then from (\ref{com1}) for the asymptotics of $\mathrm{Re}\tr G^{in}_{|m_{j}|}(R,R;iy)$ for $|y| \to \inf$ one obtains
\beq \label{com2}
\bal
&\mathrm{Re}\tr G^{in}_{|m_{j}|}(R,R;iy) \to\frac{2}{R^2|y|^3}\(\frac{m_{j}^{2}}{R}V_{0}+Q\)+\frac{6}{R^4 |y|^5}\[-\frac{m_{j}^{4}}{2R}V_{0}+\frac{m_{j}^{2}}{2R}V_{0}\right.\\
&\left. -R\(\frac{2}{3}m_{j}^{2}V_{0}^{3}+m_{j}^{2}V_{0}\)-R^{3}\(\frac{2}{3}V_{0}^{3}+\frac{V_{0}}{2}\)\]-\underbrace{-\frac{1}{2  R^3 y^{4}}\(\frac{m_{j}^2 Q}{R^{2}}+Q\)}_{\mathrm{Re}\(D_{4}^{in}(R)+D_{4}^{in}(R)|_{m_{j} \to - \m_{j}}\)/y^4}\\
&\underbrace{-\frac{2}{R^3 |y|^{5}}\(\underbrace{-\frac{m_{j}^4 Q}{4 R^3}+\frac{m_{j}^4 Q}{4 R^3}}_{=0}+\underbrace{\frac{m_{j}^2 Q}{8 R^{3}}-\frac{5 m_{j}^2 Q}{8 R^{3}}}_{=-\frac{m_{j}^2}{2 R^{2}}V_{0}}\underbrace{-\frac{m_{j}^2 Q}{4 R}+\frac{m_{j}^2 Q}{4 R}}_{=0}\underbrace{-\frac{Q}{4 R}}_{=-\frac{V_{0}}{4}}\)}_{\mathrm{Re}[\(D_{5}^{in}(R)+D_{5}^{in}(R)|_{m_{j} \to -m_{j}}\)/(i |y|^5)]}+ \mathrm{O}\(\frac{1}{|y|^{6}}\) \ .
\eal
\eeq
Now let us study the order $\mathrm{O}\(1/|y|^{5}\)$ in (\ref{com2}) in more detail
\beq \label{com3}
\bal
&\frac{6}{R^4 |y|^5}\[-\frac{m_{j}^{4}}{2R}V_{0}+\frac{m_{j}^{2}}{2 R}V_{0}-R\(\frac{2}{3}m_{j}^{2}V_{0}^{3}+m_{j}^{2}V_{0}\)-R^{3}\(\frac{2}{3}V_{0}^{3}+\frac{V_{0}}{2}\)+\frac{m_{j}^2}{6 R}V_{0}+R\frac{V_{0}}{12}\]\\
&=\frac{6}{R^4 |y|^5}\[-\frac{m_{j}^{4}}{2R}V_{0}+\frac{2m_{j}^{2}}{3 R}V_{0}-R\(\frac{2}{3}m_{j}^{2}V_{0}^{3}+m_{j}^{2}V_{0}-\frac{V_{0}}{12}\)-R^{3}\(\frac{2}{3}V_{0}^{3}+\frac{V_{0}}{2}\)\] \ .
\eal
\eeq
As a result, there follows from (\ref{com1})-(\ref{com3}) precisely the asymptotics of $\mathrm{Re} \tr G_{|m_j|}(R,R;iy)$ (\ref{c11}).

Quite analogously, the general form of the asymptotics of $\mathrm{Re} \, \tr G_{|m_{j}|}^{out}(r,r;iy)$  for $|y| \to \inf$  takes the form
\beq \label{com4}
\bal
&\mathrm{Re}\tr G^{out}_{|m_{j}|}(r,r;iy) \to\frac{2}{r^2|y|^3}\(\frac{m_{j}^{2}}{r}\frac{Q}{r}+Q\)
\\&+ \frac{6}{r^4 |y|^5}\[-\frac{m_{j}^{4}}{2r}\frac{Q}{r}+\frac{5 m_{j}^{2}}{6 r}\frac{Q}{r}-r\(\frac{2}{3}m_{j}^2 \(\frac{Q}{r}\)^3+m_{j}^2 \frac{Q}{r}-\frac{Q}{6 r}\)-r^3 \(\frac{2}{3}\(\frac{Q}{r}\)^3+\frac{Q}{2 r}\)\]\\ &
+ \mathrm{O}\(\frac{1}{|y|^{7}}\)+\mathrm{Re}\( \left.\mathrm{e}^{-2 \g \(r-R\)}\right|_{\e=iy}\(\frac{R}{r}\)^{i 2 Q}\right.\\
&\left.\times\[\frac{D^{out}_{4}(r)+D^{out}_{4}(r)|_{m_j \to -m_j}}{\e^{4}}+\frac{D^{out}_{5}(r)+D^{out}_{5}(r)|_{m_j \to - m_j}}{\e^{5}}+\mathrm{O}\(\frac{1}{|\e|^6}\)\]\) \ .
\eal
\eeq
Then from (\ref{com4}) for the asymptotics of $\mathrm{Re}\tr G^{out}_{|m_{j}|}(R,R;iy)$ for $|y| \to \inf$ there follows
\beq \label{com5}
\bal
&\mathrm{Re}\tr G^{out}_{|m_{j}|}(R,R;iy)\to \frac{2}{R^2|y|^3}\(\frac{m_{j}^{2}}{R}\underbrace{\frac{Q}{R}}_{=V_{0}}+Q\)+ \frac{6}{R^4 |y|^5}\[-\frac{m_{j}^{4}}{2R}\underbrace{\frac{Q}{R}}_{=V_{0}}+\frac{5 m_{j}^{2}}{6 R}\underbrace{\frac{Q}{R}}_{=V_{0}}\right.\\
&\left.-R\(\frac{2}{3}m_{j}^2 \(\underbrace{\frac{Q}{R}}_{=V_{0}}\)^3+m_{j}^2 \underbrace{\frac{Q}{R}}_{=V_{0}}-\underbrace{\frac{Q}{6R}}_{=\frac{V_{0}}{6}}\)-R^3 \(\frac{2}{3}\(\underbrace{\frac{Q}{R}}_{=V_{0}}\)^3+\underbrace{\frac{Q}{2R}}_{=\frac{V_{0}}{2}}\)\]\\
&\underbrace{-\frac{1}{2 R^3 y^4}\(\frac{m_{j}^2 Q}{R^2}+Q\)}_{\mathrm{Re}\(D_{4}^{out}(R)+D_{4}^{out}(R)|_{m_{j} \to - m_{j}}\)/y^4}
\eal
\eeq
\begin{equation*}
\underbrace{-\frac{2}{R^3 |y|^5}\(\underbrace{\frac{m_{j}^4 Q}{4R^3}-\frac{m_{j}^4 Q}{4 R^3}}_{=0}\underbrace{-\frac{m_{j}^2 Q}{8 R^3}+\frac{5 m_{j}^2 Q}{8 R^3}}_{\frac{ m_{j}^2}{2 R^2}V_{0}}\underbrace{+\frac{m_{j}^2 Q}{4 R}-\frac{m_{j}^2 Q}{4 R}}_{=0}\underbrace{+\frac{Q}{4 R}}_{=\frac{V_{0}}{4}}\)}_{\mathrm{Re}[\(D_{5}^{out}(R)+D_{5}^{out}(R)|_{m_{j} \to - m_{j}}\)/(iy^5)]}+ \mathrm{O}\(\frac{1}{|y|^{6}}\)\ .
\end{equation*}
Again, let us study the order $\mathrm{O}\(1/|y|^{5}\)$ in (\ref{com5}) more carefully:
\beq\label{com6}
\bal
&\frac{6}{R^4 |y|^5}\[-\frac{m_{j}^{4}}{2R}V_{0}+\frac{5 m_{j}^{2}}{6 R}V_{0}-R\(\frac{2}{3}m_{j}^2 V_{0}^{3}+m_{j}^2 V_{0}-\frac{V_{0}}{6}\)\right.\\
&\left.-R^3 \(\frac{2}{3}V_{0}^{3}+\frac{V_{0}}{2}\)-\frac{m_{j}^{2}}{6 R}V_{0}-R\frac{V_{0}}{12}\]\\
&=\frac{6}{R^4 |y|^5}\[-\frac{m_{j}^{4}}{2R}V_{0}+\frac{2m_{j}^{2}}{3 R}V_{0}-R\(\frac{2}{3}m_{j}^{2}V_{0}^{3}+m_{j}^{2}V_{0}-\frac{V_{0}}{12}\)-R^{3}\(\frac{2}{3}V_{0}^{3}+\frac{V_{0}}{2}\)\] \ ,
\eal
\eeq
whence it follows that  (\ref{com4})-(\ref{com6}) precisely reproduce the asymptotics of $\mathrm{Re} \tr G_{|m_j|}(R,R;iy)$ (\ref{c11}).

Proceeding further, by taking into account that the subtraction of $2 \, \tr G^{(1)}_{m_j}(r;iy)$  removes all the linear in $Q$ and $V_0$ terms, from (\ref{c10})-(\ref{c12}) one finds the asymptotics of $\mathrm{Re}\[\tr G_{|m_j|}(r,r;iy)-2 \, \tr G^{(1)}_{m_j}(r;iy)\]$ for $|y| \to \inf$. The main result is that the first non-vanishing term in the asymptotics of $\mathrm{Re}\[\tr G_{|m_j|}(r,r;iy)-2 \, \tr G^{(1)}_{m_j}(r;iy)\]$ turns out to be proportional to  $Q^{3}/|y|^{5}\times \text{multiplier, depending only on $m_j$ and $r$} $. The factor $Q^3$ underlines that by construction $\r^{(3+)}_{VP}(r)$ does not contain any linear in $Q$ terms, while the asymptotical behavior  $\sim |y|^{-5}$ by itself guarantees the uniform convergence of the integral $\int \! dy \ \mathrm{Re}\[ \tr G_{|m_j|}(r,r;iy)-2\, \tr G^{(1)}_{m_j}(r;i y)\]$ with respect to $m_j$ and $r$.

It would be also worth-while noticing that the asymptotics (\ref{c10})-(\ref{c12}) cannot be used in the infinitesimal vicinity of the point $r=0$, since  $r$ enters into $\tr G_{m_{j}}(r,r;iy)$ via combinations $r \sqrt{1+y^2}$ and  $ r \sqrt{1-(iy+V_0)^2}$, which for $|y| \to \inf$ in the vicinity of $r=0$ might remain finite. In the case under consideration, however, when the external potential (\ref{1.00}) is regular at the origin, there follows from the obvious physical reasons that the induced vacuum density should also be finite and continuous at the origin. Therefore it can be obtained by means of a limit transition by continuity from the region, where $r$ is non-zero. Hence, the uniform convergence of the integral $\int \! dy \ \mathrm{Re}\[ \tr G_{|m_j|}(r,r;iy)-2\, \tr G^{(1)}_{m_j}(r;i y)\]$ holds also for $r=0$.

At the next stage let us use the asymptotics of $\tr G_{m_j}(r,r;iy)$ for $r\to \infty$, substantially specified compared to (\ref{3.22}), namely
\beq\label{com7}
\bal
\tr G_{m_{j}}(r,r;iy)&\to \frac{i y}{\sqrt{1 + y^2} r} +  \frac{Q}{\(1 + y^2\)^{3/2}r^2}\\
&+ \frac{1}{2 \(1 + y^2\)^{5/2} r^{3}}\( -i m_j^2 y^3-i m_j^2 y+m_j y^2+m_j+3 i Q^2 y\) \\
&+\frac{1}{2 \(1 + y^2\)^{7/2} r^{4}}\(2 m_{j}^2 Q y^4+m_{j}^2 Q y^2-m_{j}^2 Q+3 i m_{j} Q y^3\right.\\
&\left.+3 i m_{j} Q y-4 Q^3 y^2+Q^3+Q y^2+Q\)+\mathrm{O}\(\frac{1}{r^5}\) \ .
\eal
\eeq
From  (\ref{com7}) there follows that the first non-vanishing term in the asymptotics of $\mathrm{Re}\[\tr G_{|m_j|}(r,r;iy)-2 \, \tr G^{(1)}_{m_j}(r;iy)\]$ for $r\to \infty$ should be proportional to  $Q^{3}/r^4 \times \text{multiplier, depending on $m_j$ and $y$ only} $. The factor $Q^3$ underlines once more that $\r^{(3+)}_{VP}(r)$ by construction does not contain any linear in $Q$ terms. In turn, due to the uniform convergence of the integral $\int \! dy \ \mathrm{Re}\[ \tr G_{|m_j|}(r,r;iy)-2\, \tr G^{(1)}_{m_j}(r;i y)\]$ with respect to $m_j$ and $r$, established above, the asymptotics of $\r^{(3+)}_{VP}(r)$ for $r\to \infty$ turns out to be $\sim r^{-4}$ uniformly with respect to $m_j$, what in turn provides the possibility of permutation of  summation over   $m_j$ and integration over $dr$ in (\ref{q8}).

\bibliographystyle{apsrev4-1}
\bibliography{VP2D_I_arXiv}

\end{document}